# Quantum teleportation of unknown qubit beyond Bell states formalism

**Sergey A. Podoshvedov[1] and Jaewan Kim[2]**

[1]*Laboratory of Quantum Information Processing and Quantum Computing, Institute of Natural and Exact Sciences, South Ural State University (SUSU), Lenin Av. 76, Chelyabinsk, Russia*
[2]*School of Computational Sciences, Korea Institute for Advanced Study, Seoul, South Korea*

e-mail: sapodo68@gmail.com

We suggest implementation of quantum teleportation protocol of unknown qubit beyond Bell states formalism. Hybrid entangled state composed of coherent components that belong to Alice and dual-rail single photon at Bob's disposal is used. Nonlinear effect on the teleported state is realized due to peculiarity of interaction of coherent components with discrete variable state on a beam splitter. Bob performs unitary transformation after receiving the appropriate classical information. The protocol is nearly deterministic but the output state is subject to amplitude distortion which reduces the fidelity of the protocol. Final problem is reduced to finding a way to get rid of amplitude-distorting factor to restore the original qubit. Some strategies to increase the fidelity of the protocol are used. We show the teleportation can be implemented with arbitrary settings. Our approach is applicable to different basic number states of original unknown qubit. The scheme is realized by linear optics methods and requires an irreducible number of optical elements.

In 1993, a seminal paper[1] presented a quantum information protocol that became one of the most interesting and exiting manifestation of the quantum nature of physical objects. The process requires the resource of quantum entanglement[2-3]. Without its presence, quantum state transfer is not possible within the laws of quantum mechanics. In this protocol, an unknown quantum state of a physical system is measured together with part of entangled state (or quantum channel) and subsequently reconstructed at a remote location while the physical constituents of the original system remain at one place. Two bits of classical information are required to restore the original qubit. The transmission of unknown discrete variable (DV) qubit through space excludes superluminal communication. Quantum teleportation plays an active role in progress of quantum information[4]. The protocol itself is conceptual and makes a crucial contribution to the development of formal quantum informatics. It also represents a fundamental ingredient to development of many quantum technologies. So two-qubit gates can be realized by simultaneous teleportation of two qubits[5]. Quantum repeaters[6] are proposed for quantum communications over large distances. Gate teleportation strategies lie at the core of cluster state quantum computing[7]. A number of qubits are prepared in a multi-partite entangled state (e.g., a cluster state). Then, a corresponding measurement of one qubit teleports its state onto another particle on which the desired quantum gate operates[7,8]. Quantum teleportation protocol is the basis for building a quantum computer[9-11] and plays an important role to implement it fault-tolerantly. The ability to teleport an unknown qubit is confirmed in practice[12,13]. The implementation of two-qubit operations is experimentally shown[14].

Quantum teleportation is not restricted to the DV qubits, but it can be implemented in higher-dimensional systems, in particular, it can be extended to quantum systems in an infinite-dimensional Hilbert space, known as continuous-variable (CV) systems[15]. Such type of quantum teleportation is typically realized in position- and momentum-like quadrature representation[16]. In such interpretation, the entangled resource corresponds to a two-mode squeezed vacuum state[17].

CV teleportation is realized in deterministic manner but its fidelity cannot reach unit as CV resource does not provide a maximally entangled state. In the same time, the practical realization of complete Bell states detection being main ingredient of the DV quantum teleportation is major problem for its deterministic implementation since only two Bell states are discriminated by linear optics and photodetection, thus limiting the protocol to 50% efficiency[18] but with perfect fidelity. In principle, using entangled quantum channel with large number $n$ of qubits, one can teleport one unknown qubit with better efficiency[10] approaching 100% for infinite $n$. But this theoretical possibility can hardly be realized at least at the current level of technological development. Now, improving the practical efficiency of the Bell states detection is active area of modern investigations[19-22]. In order to overcome the limitation, use of entangled states composed of DV and CV ones can become promising[23,24]. Such an approach is aimed to make use of advantages of DV and CV states to teleport unknown qubit with larger success probability and high fidelity. Implementations of the hybrid entanglement between a coherent qubit (superposition of coherent states (SCS)) and microscopic qubit of vacuum and single photon[25,26] were demonstrated. The implementation of the hybrid state between coherent components and dual-rail single photon is proposed[27]. The generation of a more complex type of hybrid entanglement was reported[28].

Here, we present a novel implementation of quantum teleportation of an unknown qubit. The hybrid entangled state composed of coherent components with equal modulo but different in sign amplitudes and dual-rail single photon is used. The teleported unknown qubit is DV state. Nonlinear effect on the target state in Bob's hands is realized due to peculiarities of interaction of CV and DV states on BS[29-31] (DV-CV interaction). Alice recognizes all her measurement outcomes as in CV teleportation and Bob, after receiving additional classical information, gets at his disposal set of the states that are subject to controlled amplitude distortion. Efficient implementation of the quantum teleportation is reduced to the procedure of getting rid of the known amplitude factors to obtain original unknown qubit. The implementation is realized beyond Bell states formalism and has some similarity with CV teleportation. We are also going to call it DV-CV quantum teleportation of unknown qubit. Strategies aimed at improving the effectiveness of the protocol in terms of the fidelity of output states are considered. We show DV-CV teleportation can be implemented with arbitrary settings. We show that this approach is applicable to the qubits composed of arbitrary base number states. Developed approach is based on the use of the properties of the displaced number states, the interest to which begins to revive[32-38]. Various aspects of generation and nonclassical properties of the experimentally observed displaced states were discussed[32-35]. The displaced states have an additional degree of freedom of the displacement amplitude that extends the possibilities for manipulating them to generate quantum gates[36,37] and also for generating of SCS of large amplitude[38]. SCS with the displacement amplitude sufficient for the implementation of the protocol is experimentally generated[39].

## Results
**Simplified version of the DV-CV teleportation.** Suppose Alice has at her disposal unknown qubit

$$|\varphi^{(01)}\rangle_2 = a_0|0\rangle_2 + a_1|1\rangle_2, \qquad (1)$$

satisfying the normalization condition $|a_0|^2 + |a_1|^2 = 1$, where subscripts denote the state modes. Explanation of the used superscript is given in section Methods. Alice is going to transmit unknown qubit to Bob which is located at a considerable distance apart from Alice. Alice cannot directly send him this qubit but she has at her disposal a part of the hybrid entangled state

$$|\Psi\rangle_{134} = (|0,-\beta\rangle_1|01\rangle_{34} + |0,\beta\rangle_1|10\rangle_{34})/\sqrt{2}, \qquad (2)$$

which was created beforehand. Here, the notation for the displaced number states given by Eq. (57) is used. The amplitude $\beta$ of the state is assumed to be positive $\beta > 0$. The entangled hybrid state in Eq. (2) consists of coherent components with opposite in sign amplitudes and dual-rail single photon. The coherent mode (mode 1) belongs to Alice, while single photon which simultaneously take two modes (modes 3 and 4) is at Bob's disposal. Note the unknown qubit in Eq. (1) is defined in the basis $\{|0\rangle, |1\rangle\}$, while Bob's photon is determined in the base $\{|01\rangle, |10\rangle\}$. Alice performs a measurement of her observable. In order to perform the measurement, the unknown state and the coherent components of the state in Eq. (2) are mixed on beam splitter shown in Fig. 1(a)

$$BS_{12} = \begin{bmatrix} t & -r \\ r & t \end{bmatrix}, \quad (3)$$

where the real parameters $t > 0, r > 0$ are the transmittance $t$ and the reflectance $r$, respectively, satisfying the normalization condition $t^2 + r^2 = 1$. Here, subscripts 12 concerns modes of the mixed states. The interaction is example of DV-CV interaction on beam splitter.

Before considering the general case, consider simplified version of the DV-CV interaction in the case of $\beta \ll 1$. Then, the coherent components of the state in Eq. (2) can be written

$$|0, -\beta\rangle \approx |0\rangle - \beta|1\rangle, \quad (4a)$$
$$|0, \beta\rangle \approx |0\rangle + \beta|1\rangle. \quad (4b)$$

The hybrid entangled state in Eq. (2) can be rewritten

$$|\Psi\rangle_{134} \approx (|0\rangle_1 - \beta|1\rangle_1)|01\rangle_{34} + (|0\rangle_1 + \beta|1\rangle_1)|10\rangle_{34}, \quad (5)$$

where we leave out the normalization factor. The teleported single photon may pass through the beam splitter simultaneously with the reflected one from the state in Eq. (5) so that all information about where the photon came from is erased. APD located at second mode may register the coming photon (registration of the event $|01\rangle_{12}$) that generates the following non normalized state at Bob's location

$$(ta_1 + \beta r a_0)|01\rangle_{34} + (ta_1 - \beta r a_0)|10\rangle_{34}. \quad (6a)$$

If photon of the quantum channel transmits through the beam splitter while the photon of the teleported qubit in Eq. (1) is reflected with erasure of all information which event has happened, then Bob obtains the following state

$$(ra_1 - \beta t a_0)|01\rangle_{34} + (ra_1 + \beta t a_0)|10\rangle_{34}, \quad (6b)$$

provided that Alice has measured the outcome $|10\rangle_{12}$. Alice encodes her measurement results by two bits either (01) or (10) and sends the message to Bob so that he can identify the obtained states. After Bob receives this information, he does $P(\pi)$ conversion on one of his, for example, on mode 3 acting as $P(\pi)|0\rangle \to |1\rangle$, $P(\pi)|1\rangle \to -|1\rangle$ followed by Hadamard transformation on dual-rail single photon

$$H|01\rangle \to (|01\rangle + |10\rangle)/\sqrt{2}, \quad (7a)$$
$$H|10\rangle \to (|01\rangle - |10\rangle)/\sqrt{2}, \quad (7b)$$

that leaves him with the states

$$a_0|01\rangle_{34} + B_{01} a_1|10\rangle_{34}, \quad (8a)$$
$$a_0|01\rangle_{34} + B_{10} a_1|10\rangle_{34}, \quad (8b)$$

where additional amplitude-distorting factors are determined by

$$B_{01} = 1/\alpha, \quad (9a)$$
$$B_{10} = r/(\beta t), \quad (9b)$$
$$\alpha = \beta r/t. \quad (9c)$$

Selecting the appropriate values of the used parameters (either $B_{01} = 1/\alpha = 1$ with $\alpha = 1$ or $B_{10} = 1$) allows him to restore original qubit in the base $\{|01\rangle, |10\rangle\}$. Note the Hadamard operation on a single photon can be implemented using the balanced beam splitter and phase shift operations[40].

This simplified consideration is an example of a new type of quantum teleportation of an unknown qubit different form DV[1] and CV[15] teleportation. The Bell state measurement is not used and the protocol under study is performed beyond the Bell states formalism. CV quantum teleportation of an unknown qubit is accomplished using a two-mode squeezed vacuum, which has perfect correlating properties when the squeezing parameter tends to infinity. Otherwise, the imperfect CV entanglement teleports an unknown qubit only with some fidelity. Hybrid state with opposite in sign displacement amplitudes is used in our case. Moreover, different mechanism of interaction is employed. Therefore, we are going to call it DV-CV teleportation of unknown qubit.

**From simplified model to real consideration of DV-CV teleportation.** In the previous section, we proved the feasibility of the DV-CV mechanism in a simplified consideration. Now, we consider general case. The same optical scheme in Fig. 1(a) as in previous section is used. Alice mixes the coherent components of the entangled state in Eqs. (2), (78) with unknown qubit in Eq. (1), (77) on the beam splitter followed by parity recording in coherent mode (mode 1) and defining the number of photons $m$ in the teleported mode (mode 2) as shown in Fig. 1(a). Corresponding mathematical apparatus is presented in section Methods. In the Methods section, we prove the feasibility of the mechanism with arbitrary settings (the amplitude $\beta$ of the state in Eq. (2) and the transmittance $t$ and the reflectance $r$ of the BS, respectively, in Eq. (3)) on example of the quantum teleportation of unknown qubit in Eq. (1). We use the case of highly transmissive beam splitter (HTBS) with transmittance $t \to 1$ and the reflectance $r \to 0$ but arbitrary amplitude $\beta$ of the hybrid state in Eq (2) in our consideration. Given approach is generalized on the quantum teleportation of an unknown state in Eq. (77) with help of the state in Eq. (78) with arbitrary amplitude $\beta$.

Coherent components of the hybrid state in Eqs. (2), (78) simultaneously displace the unknown qubit in Eqs. (1), (77) in indistinguishable manner by the different displacement values. All information about value of the displacement disappears. Measurement of the original state $BS_{12}\left(|\Psi_\phi\rangle_{134}|\varphi^{(kn)}\rangle_2\right)$ collapses it onto another state at Bob's disposal subject to controlled$-Z$ operation. Result of the parity measurement $j$ in coherent mode is coded by the values 0 and 1 ($j = 0,1$), while the number of measured photons can take any natural values $m$. Forwarding her measurement outcomes to Bob, Alice helps him to create original states, in general case, subject to amplitude distortion. To obtain amplitude-modulated states Bob must apply Pauli single-qubit transformations $\left(Zexp(i\phi/2)R_Z(\phi)\right)^{k-m}$ removable by action of Hadamard gate $H$, where $Z-$Pauli matrix define transformation $R_Z(\phi)$ being a rotation operation of qubit about $Z$ axis. These operations are easily implemented by linear optics devices on single photon[38]. If Bob successfully demodulates an amplitude-modulated unknown qubit, then we can talk about quantum teleportation with perfect fidelity. The terminology used (amplitude-modulated qubit, demodulation) is introduced in the Methods section. We note that Bob can completely identify all obtained states, in contrast to Bell states treatment[18] (the protocol beyond Bell states formalism). As in the case of CV teleportation of an unknown qubit, all measurement outcomes in DV-CV teleportation are entirely distinguishable (deterministic operation) but the fidelity of the teleported qubit becomes imperfect.

Matrix elements[30,31] $c_{km}(\alpha)$ in Eq. (59) determine the amplitude-distorting factors as well as the success probability of success of an event. Consider some of them. So, we have for the coherent state $|0,\alpha\rangle$

$$c_{0m}(\alpha) = \alpha^m/\sqrt{m!}, \qquad (10)$$

for the displaced singe photon $|1,\alpha\rangle$

$$c_{1m}(\alpha) = \alpha^{m-1}(m - |\alpha|^2)/\sqrt{m!}, \qquad (11)$$

for the displaced two-photon state $|2,\alpha\rangle$

$$c_{2m}(\alpha) = \alpha^{m-2}(m(m-1) - 2m|\alpha|^2 + |\alpha|^4)/(\sqrt{2}\sqrt{m!}), \qquad (12)$$

for the displaced three-photon state $|3,\alpha\rangle$

$$c_{3m}(\alpha) = \alpha^{m-3}(m(m-1)(m-2) - 3m(m-1)|\alpha|^2 + 3m|\alpha|^4 - |\alpha|^6)/(\sqrt{3!}\sqrt{m!}). \qquad (13)$$

Using the expressions, we can construct amplitude-distorting factors $A_m^{(kn)}$ in Eq. (84) for some values of $k$ and $n$ in the case of use of HTBS

$$A_m^{(01)}(\alpha) = \frac{m-|\alpha|^2}{\alpha}, \qquad (14)$$

$$A_m^{(02)}(\alpha) = \frac{m(m-1)-2m|\alpha|^2+|\alpha|^4}{\sqrt{2!}\alpha^2}, \tag{15}$$

$$A_m^{(03)}(\alpha) = \frac{m(m-1)(m-2)-3m(m-1)|\alpha|^2+3m|\alpha|^4-|\alpha|^6}{\sqrt{3!}\alpha^3}, \tag{16}$$

$$A_m^{(12)}(\alpha) = \frac{m(m-1)-2m|\alpha|^2+|\alpha|^4}{\sqrt{2!}\alpha(m-|\alpha|^2)}. \tag{17}$$

Mathematical details of realization of controlled−$Z$ gate for the DV state at Bob's qubit are presented in section Methods. In particular, we have the following relation $c_{2mn}(-\alpha) = (-1)^n c_{2mn}(\alpha)$ for even $l = 2m$ displaced number states and $c_{2m+1n}(-\alpha) = (-1)^{n-1} c_{2m+1n}(\alpha)$ for odd $l = 2m + 1$ displaced number states. So, the realization of the nonlinear effect in DV-CV interaction for unknown qubit in Eq. (1) is ensured by the property of matrix elements $c_{0n}(-\alpha) = (-1)^n c_{0n}(\alpha)$ and $c_{1n}(-\alpha) = (-1)^{n-1} c_{1n}(\alpha)$ to change their sign in dependency on parity of the basic state when changing the displacement amplitude on opposite $\alpha \to -\alpha$. The amplitude-distorting factors $A_m^{(kn)}$ in Eqs. (14-17) and Eq. (84) arise as a result of the fact that the displaced number states are transformed differently when projecting them onto measurement basis of the number states. The presence of this additional factor $A_m^{(kn)}$ is a distinctive feature inherent to the DV-CV interaction. One can even say that the CV state leaves its imprint in the teleported DV state. Note that the value of only the amplitude-distorting factor $A_m^{(kn)}$ is familiar to the participants of the protocol as it can be calculated using, for example, the formulas (14-17). It is believed that Alice and Bob know exactly the parameters $\beta$, $t$ and $r$. The phase of the unknown qubit is not affected by the distorting factor.

Consider implementation of the DV-CV teleportation of the unknown state in Eq. (1) with help of the state in Eq. (2). The success probability $P_m^{(01)}$, obtained by summing over all measurement outcomes in the first "coherent" mode, for Bob to have AM unknown qubit (the term is introduced in section Methods) follows from Eq. (75)

$$P_m^{(01)} = exp(-|\alpha|^2) \frac{|\alpha|^{2n}\left(1+\left(\left|A_m^{(01)}\right|^2-1\right)|a_1|^2\right)}{n!}, \tag{18}$$

It can be directly checked the success probabilities satisfy the normalization condition $\sum_{m=0}^{\infty} P_m^{(01)} = 1$. It should be noted that the success probability of an event becomes dependent on an unknown parameter $|a_1|$. This dependence is associated with the appearance of the amplitude-distorting factor $A_m^{(kn)}$ in the teleported qubit. In general, we have an infinite number of measurement outcomes. But since the parameter $\beta$ in Eq. (2) is free to change, we can achieve such a situation when success probabilities of certain measurement outcomes will significantly exceed all remaining. Indeed, numerical results in Fig. 2(a) show that the condition $P_0^{(01)} + P_1^{(01)} \approx 1$ is performed in the case of $\alpha < 1$ with high accuracy. We only note that in the case of small values of $\beta < 1$, we can make use of the formulas (89, 90). Then, Alice's measurement in the base $\{|00\rangle, |01\rangle, |10\rangle, |11\rangle\}$ instead of $\{|even\rangle_1|0\rangle_2, |even\rangle_1|1\rangle_2, |odd\rangle_1|0\rangle_2, |0dd\rangle_1|1\rangle_2\}$ ($|even\rangle \approx |0\rangle$, $|odd\rangle \approx |1\rangle$ for $\beta < 1$) with the subsequent transfer of two bits of information $\{(00),(01),(10),(11)\}$ to Bob over a standard communication channel allows him to create one of two possible states either $|\Psi_0^{(01)}\rangle$ or $|\Psi_1^{(01)}\rangle$ (the relevant results of the protocol are also presented in Table 1) in modes 3 and 4

$$|\Psi_0^{(01)}\rangle = N_0^{(01)} \begin{bmatrix} a_0 \\ a_1 A_0^{(01)} \end{bmatrix}, \tag{19}$$

$$|\Psi_1^{(01)}\rangle = N_1^{(01)} \begin{bmatrix} a_0 \\ a_1 A_1^{(01)} \end{bmatrix}, \tag{20}$$

with the normalization factor $N_m^{(01)} = \left(1 + \left(\left|A_m^{(01)} - 1\right|^2\right)|a_1|^2\right)^{-0.5}$. Bob obtains the states after application of unitary operations $HZ^r$ with $r = j + m$, where $j$ is the parity (even or odd) of the state in "coherent" first mode. Notice the simplified model may also give similar results in partial case of the measurement outcomes $\{|01\rangle, |10\rangle\}$. It is worth noting the coefficient $B_{01}$ in Eq. (9a) almost coincides modulo with $A_0^{(01)}$ in Eq. (14) $\left(B_{01} \approx A_0^{(01)}\right)$ in the case of $\alpha < 1$.

The success probabilities $P_m^{(kn)}$ for different values $k$ and $n$ can be calculated using Eq. (92). Numerical dependencies of the success probabilities $P_m^{(02)}$, $P_m^{(03)}$, and $P_m^{(12)}$ on amplitude of the unknown qubit $|a_1|$ for different values of the displacement amplitudes $\alpha$ are shown in Figs. 2(b), 2(c) and 2(d), respectively. Numerical observations show that, at least for the parameters $k$ and $n$ used, takes place relation

$$P_k^{(kn)} + P_n^{(kn)} \approx 1, \qquad (21)$$

in the case of $\alpha < 1$. The relation in equation (21) means that contribution of the events with $m = k$ and $m = n$ prevails over all other events. All other events with $m \neq k$ and $m \neq n$ can be neglected since their contribution is negligible. In fact, this relationship directly follows from the definition of matrix elements in Eq. (59). In the case, Alice must send Bob two bits of classical information to help him to create AM unknown qubit either $|\Psi_k^{(kn)}\rangle$ or $|\Psi_n^{(kn)}\rangle$, respectively, given in Eqs. (82).

| Measurement outcomes | Teleported states | Success probabilities |
|---|---|---|
| $(0, k), (1, k)$ | $|\Psi_k^{(kn)}\rangle$ (AM) | $P_k^{(kn)}$ |
| $(0, n), (1, n)$ | $|\Psi_n^{(kn)}\rangle$ (AM) | $P_n^{(kn)}$ |

**Table 1**. Realization of quantum teleportation protocol of the unknown qubit in Eq. (77) with help of hybrid entangled state in Eq. (78) in the case of $\alpha < 1$, when Bob obtains amplitude modulated versions of the original qubit either $|\Psi_k^{(kn)}\rangle$ or $|\Psi_n^{(kn)}\rangle$ in Eqs. (19) and (20) with success probability $P_k^{(kn)} + P_n^{(kn)} \approx 1$.

It is interesting to consider what state will be received by Bob after Alice has performed the measurement but before Bob has learned the measurement results. We only do corresponding calculations on example of the unknown qubit in Eq. (1). After Alice measurement, Bob obtains the state $\varrho_B^{(01)}$ described by the density matrix

$$\varrho_B^{(01)} = \sum_{k=0}^{\infty} \sum_{p=0}^{\infty} P_{pk}^{(01)} |\varphi_{pk}^{(01)}\rangle \langle \varphi_{pk}^{(01)}|, \qquad (22)$$

where the expression for $P_{pk}^{(01)}$ is taken from Eq. (75), while the state $|\varphi_{pk}^{(01)}\rangle$ is given by Eq. (71). Doing the same calculations as we did for formula output in Eq. (76), one obtains final Bob's state after Alice's measurement

$$\varrho_B^{(01)} = \frac{1}{2}(|01\rangle\langle 01| + |10\rangle\langle 10|). \qquad (23)$$

Thus, the obtained state has no dependence on parameters of the teleported state preventing Alice from using the teleportation technique to transmit information to Bob faster than speed of light[1,4]. Note the hybrid state in Eq. (2) is not maximally entangled. The negativity $\mathcal{N}$ of a bipartite composed system in Eq. (2) is equal to $\mathcal{N} = \sqrt{1 - exp(-4|\beta|^2)}$. It becomes maximally entangled $\mathcal{N}_{max} = 1$ only in the case of its amplitude $\beta$ approaching to infinity $\beta \to \infty$.

**Quantum teleportation of initially AM unknown qubit.** We have shown the use of small values of the displacement amplitude $\alpha$ on which we need to displace the unknown qubit allows us significantly to decrease (2 bits) the amount of classical information sent to recipient.

Parity measurement and photon number resolving measurement can be replaced by on-off measurement that can be realized by commercially achievable avalanche photodiode (APD). Small amplitude $\beta$ of the coherent components of the hybrid state in Eq. (78) can be used that gives an additional benefit to the implementation of the quantum teleportation of unknown qubit since such states will actually be implemented in practice. Nevertheless, the problem of amplitude demodulation, or the same getting rid of additional controlled by Alice distortion factors, of the output qubits remains which impairs the fidelity of the teleportation.

Instead of considering demodulation methods applicable to the teleported qubit considered in previous section, we consider quantum teleportation of initially AM unknown qubit in Fig. 1(b). Suppose that the third party of the protocol (Victor) modulates the unknown qubits and transmits them to Alice. The amplitudes $a_0$ and $a_1$ remain unknown but the states acquire an additional amplitude factors determined by Victor but known to Alice and Bob. Aim of Alice and Bob is to teleport the initially AM unknown qubit and send to Viktor the original unknown qubit without amplitude distortion for verification. We are going to consider two types of initially AM unknown qubits prepared by Victor

$$|\varphi_{AMk}^{(kn)}\rangle = N_{AMk}^{(kn)} \begin{bmatrix} a_0 \\ A_k^{(kn)-1} a_1 \end{bmatrix}, \tag{24}$$

$$|\varphi_{AMn}^{(kn)}\rangle = N_{AMn}^{(kn)} \begin{bmatrix} a_0 \\ A_n^{(kn)-1} a_1 \end{bmatrix}, \tag{25}$$

where the normalization factors are given by

$$N_{AMk}^{(kn)} = \left(1 + \left(\left|A_k^{(kn)}\right|^2 - 1\right)|a_1|^2\right)^{-0.5}, \tag{26}$$

$$N_{AMn}^{(kn)} = \left(1 + \left(\left|A_n^{(kn)}\right|^2 - 1\right)|a_1|^2\right)^{-0.5}. \tag{27}$$

The choice of the initial AM states is caused by that to provide the greatest possible success probability for Bob to obtain the original qubit not AM one. The same HTBS is used to mix coherent components of the hybrid state in Eq. (78) with unknown qubit in Eq. (77) with subsequent parity measurement in the first mode and the number of photons in the second mode. Again, Alice does measurement in the base $\{|00\rangle, |01\rangle, |10\rangle, |11\rangle\}$ for $\beta < 1$. In optical scheme in Fig. 1(b), Victor generates AM unknown qubits in Eqs (24), (25). Bob's efforts to get rid of amplitude-distorting factors for his qubits are shown in the optical scheme in Figure 1(b).

The same DV-CV interaction mechanism described in the previous section is used. When Alice forwards some amount of classical information, Bob can also apply sequence of unitary operations shown in Fig. 1(b) to his photon. Finally, Bob obtains the AM states. The output states involve amplitude-distorting terms either $A_k^{(kn)-1} A_m^{(kn)}$ for the input state in Eq. (24) or $A_n^{(kn)-1} A_m^{(kn)}$ for the input state in Eq. (25) provided that Alice defined parity of CV state in first mode and registered $m$ photons in second mode. If the number of measured photons coincides with either number $k$ ($m = k$) for the input state in Eq. (24) or number $n$ ($m = n$) for the input state in Eq. (25), then Bob obtains original unknown qubit $\left(A_k^{(kn)-1} A_k^{(kn)} = A_n^{(kn)-1} A_n^{(kn)} = 1\right)$ in Eq. (77) which he can submit to Victor for verification. Note only unknown qubit at Bob's disposal lives in Hilbert space with the base $\{|01\rangle, |10\rangle\}$ unlike original one in the base $\{|0\rangle, |1\rangle\}$.

If Victor sends to Alice AM state in Eq. (24), then Bob recovers the original state in the base $\{|01\rangle, |10\rangle\}$ provided that Alice measured $m = k$ photons

$$|\psi^{(kn)}\rangle = \begin{bmatrix} a_0 \\ a_1 \end{bmatrix}. \tag{28}$$

We use the notation $\psi$ instead of $\varphi$ to show that Bob's output state is defined in basis different from the initial. If Victor sends to Alice AM state in Eq. (25), then Bob again obtains the original state in Eq. (28) after Alice registered $m = n$ photons. The success probabilities for Bob to obtain original qubit in Eq. (28) are the following

$$P_{t_1}^{(kn)} = F^2 |c_{kk}|^2 N_{AMk}^{(kn)2}, \tag{29}$$

if Victor sends the input AM state in Eq. (24) to Alice and
$$P_{t_2}^{(kn)} = F^2 |c_{kn}|^2 N_{AMn}^{(kn)2}, \qquad (30)$$
if Victor sends the input AM state in Eq. (25) to Alice. In addition to the original qubit, in general case, Bob receives the following AM states
$$|\psi_{km}^{(kn)}\rangle = N_{km}^{(kn)\prime} \begin{bmatrix} a_0 \\ A_m^{(kn)} A_k^{(kn)-1} a_1 \end{bmatrix}, \qquad (31)$$
if Alice teleported the state in Eq. (24) and $m \neq k$ and
$$|\psi_{nm}^{(kn)}\rangle = N_{nm}^{(kn)\prime} \begin{bmatrix} a_0 \\ A_m^{(kn)} A_n^{(kn)-1} a_1 \end{bmatrix}, \qquad (32)$$
if Alice teleported the state in Eq. (25) and $m \neq n$, where normalization factors are the following
$$N_{km}^{(kn)\prime} = \left(1 + \left(\left|A_m^{(kn)}\right|^2 \left|A_k^{(kn)}\right|^{-2} - 1\right)|a_1|^2\right)^{-0.5}, \qquad (33)$$
$$N_{nm}^{(kn)\prime} = \left(1 + \left(\left|A_m^{(kn)}\right|^2 \left|A_n^{(kn)}\right|^{-2} - 1\right)|a_1|^2\right)^{-0.5}. \qquad (34)$$
The success probabilities for Bob to obtain AM states in Eqs. (31) and (32) are the following
$$P_{km}^{(kn)} = \frac{F^2 |c_{km}|^2 N_{AMk}^{(kn)2}}{N_{km}^{(kn)\prime 2}}, \qquad (35)$$
$$P_{nm}^{(kn)} = \frac{F^2 |c_{km}|^2 N_{AMn}^{(kn)2}}{N_{nm}^{(kn)\prime 2}}. \qquad (36)$$
The expressions (35) and (36) are applicable to the case of $n - k$ being even in the case of $\beta < 1$ due to incomplete overlapping of the states in Eqs. (85) and (86). Prepared AM states in Eqs. (24) and (25) contain normalization factors with parameter $|a_1|$ of unknown qubit. Therefore, the probabilities depend on the parameter $|a_1|$.

Consider more practical case of $\alpha < 1$. Then the number of events giving a significant contribution to the success probability is reduced to two. Numerical analysis shows the conditions
$$P_{t_1}^{(kn)} + P_{kn}^{(kn)} \approx 1, \qquad (37)$$
$$P_{t_2}^{(kn)} + P_{nk}^{(kn)} \approx 1, \qquad (38)$$
are performed in the case of $\alpha < 1$. In the case, Alice sends Bob only two bits of classic information. Increase of the parameter $\alpha$ leads to violation of conditions in Eqs. (37) and (38) $P_{t_1}^{(kn)} + P_{kn}^{(kn)} < 1$, $P_{t_2}^{(kn)} + P_{nk}^{(kn)} < 1$, respectively, and entails increase of contribution of other events with probabilities $P_{km}^{(kn)}$ with $m \neq n$ and $P_{nm}^{(kn)}$ with $m \neq k$. Plots in Figs. 3 show dependence both of the success probabilities $P_{t_1}^{(01)}$ (Fig. 3(a)), $P_{t_2}^{(01)}$ (Fig. 3(b)), $P_{t_1}^{(12)}$ (Fig. 3(c)) and $P_{t_2}^{(12)}$ (Fig. 3(d)), respectively, for Bob to obtain the original qubit in Eq. (28) (either qubit $|\psi^{(01)}\rangle$ or $|\psi^{(12)}\rangle$) and the probabilities $P_{01}^{(01)}$, $P_{02}^{(01)}$ (Fig. 3(a)), $P_{10}^{(01)}$, $P_{12}^{(01)}$ (Fig. 3(b)) for the case of $k = 0$ and $n = 1$ and $P_{12}^{(12)}$, $P_{13}^{(12)}$, (Fig. 3(c)) $P_{21}^{(12)}$, $P_{20}^{(12)}$ (Fig. 3(d)) for the case of $k = 1$ and $n = 2$ for Bob to obtain AM qubits in Eqs. (31) and (32), respectively, in dependence on $|a_1|$ for different amplitudes $\alpha$. It can be seen from the plots, if the teleported qubit is highly unbalanced with $|a_1| \ll |a_0|$ (Figs. 3(a), Fig. 3(c)) or $|a_0| \ll |a_1|$ (Figs. 3(b), 3(d)) then the success probability to teleport AM unknown qubit becomes close to one. We note that the initial amplitude modulation in Eqs. (24) and (25) with the original qubit in Eq. (77) is made by a third person (Victor). Victor's actions are considered to be preparatory and may be probabilistic. In such an examination, the probability of success of getting the AM qubit is not included in the calculation of the success probability of quantum teleportation of the AM state. Finally, the results of the quantum teleportation of initially amplitude modulated unknown qubits are presented in Tables 2 and 3 in the case of $\alpha < 1$.

| Measurement outcomes | Obtained states | Success probabilities |
|---|---|---|
| $(0,k)$, $(1,k)$ | $|\psi^{(kn)}\rangle$ (original) | $P_{t_1}^{(kn)}$ |
| $(0,n)$, $(1,n)$ | $|\psi_{kn}^{(kn)}\rangle$ (AM) | $P_{kn}^{(kn)}$ |

**Table 2**. Implementation of quantum teleportation protocol of initial AM unknown qubit in Eq. (24) in the case of $\alpha < 1$. Success probability to restore the original qubit is $P_{t_1}^{(kn)}$. $P_{kn}^{(kn)}$ is the probability for Bob to obtain AM state. The relation $P_{t_1}^{(kn)} + P_{kn}^{(kn)} \approx 1$ is performed in the case of $\alpha < 1$.

| Measurement outcomes | Obtained states | Success probabilities |
|---|---|---|
| $(0,k)$, $(1,k)$ | $|\psi_{nk}^{(kn)}\rangle$ (AM) | $P_{nk}^{(kn)}$ |
| $(0,n)$, $(1,n)$ | $|\psi^{(kn)}\rangle$ (original) | $P_{t_2}^{(kn)}$ |

**Table 3**. Implementation of quantum teleportation protocol of initial AM unknown qubit in Eq. (25) in the case of $\alpha < 1$. Success probability to restore the original qubit is $P_{t_2}^{(kn)}$. $P_{nk}^{(kn)}(\alpha)$ is the probability for Bob to obtain AM state. The relation $P_{t_2}^{(kn)} + P_{nk}^{(kn)} \approx 1$ is performed in the case of $\alpha < 1$.

**Amplitude demodulation of unknown qubits.** In the previous section, we showed that the method of initial amplitude modulation of an unknown original qubit by Victor gives a possibility to increase the efficiency of the quantum teleportation but only for the unbalanced original qubit. Therefore, the effectiveness of the protocol also depends on Bob's efforts to demodulate AM unknown qubit. So, the success probabilities for Bob to obtain original qubit can be calculated as

$$P_t = p_1 P_k^{(kn)} + p_2 P_n^{(kn)}, \qquad (39)$$

in the case of direct realization of the quantum teleportation in Fig. 1(a) without use of initially AM unknown qubits with $\alpha < 1$, where $p_1$ and $p_2$ are the success probabilities to demodulate AM qubits. If we consider quantum teleportation of AM unknown qubit, then success probability is determined by

$$P_{t_1} = P_{t_1}^{(kn)} + p_k P_{kn}^{(kn)}, \qquad (40)$$
$$P_{t_2} = P_{t_2}^{(kn)} + p_n P_{nk}^{(kn)}, \qquad (41)$$

where $p_k$ and $p_n$ are the success probabilities to demodulate AM qubits in Eqs. (31) and (32).

Consider demodulation of AM qubits on example of the qubits in Eq. (31) and (32) with $k = 0$ and $n = 1$. Consider two possible ways to demodulate AM qubits: by means of its interaction with strong coherent state on HTBS and quantum swapping procedure. Suppose Bob has in his hands the AM state $|\psi_{01}^{(01)}\rangle$ in Eq. (31). Then, he displaces the state in one of the modes by quantity $\gamma_1$ with the subsequent measurement of photons in this mode. The generated superposition contains an additional amplitude factor which, under certain conditions, may compensate for the amplitude-distorting factor $A_1^{(01)} A_0^{(01)-1}$. In the case of total compensation, Bob stays with original unknown qubit in Eq. (1). The corresponding mathematical apparatus for demodulation by means of interaction of AM qubit with strong coherent state is presented in the section Methods. Using Eqs. (40) and (41) and summing up all the success probabilities, finally, we obtain the overall success probability for Bob to obtain original unknown qubit

$$P_{t_1}^{(C)} = \frac{exp(-|\alpha|^2)}{1+\left(\left|A_0^{(01)}\right|^{-2}-1\right)|a_1|^2}\left(1 + exp(-|\gamma_1|^2)|\alpha|^2((1-|\gamma_1|^2)^2 + |\gamma_1|^2)\right), \qquad (42)$$

if AM state in Eq. (24) is used. The same approach is applicable to the AM state $|\psi_{10}^{(01)}\rangle$ to compensate distorting factor $A_0^{(01)}A_1^{(01)-1}$. Finally, we have overall success probability

$$P_{t_2}^{(C)} = \frac{exp(-|\alpha|^2)|\alpha|^2}{1+\left(\left|A_1^{(01)}\right|^{-2}-1\right)|a_1|^2}\left(1 + exp(-|\gamma_2|^2)\,|\gamma_2|^2/|\alpha|^2\right), \qquad (43)$$

if the state in Eq. (25) is handed to Alice. Here, the superscript $C$ means that Bob made use of method of interaction of AM unknown qubit with strong coherent state on HTBS. The parameters $\gamma_1$ and $\gamma_2$ are determined from relations $((1-|\alpha|^2)/\alpha^2)\gamma_1 = 1 - |\gamma_1|^2$ and $\alpha^2/(1-|\alpha|^2) = \gamma_2$. They follow from Eqs. (93) and (95), respectively. In the limit case of $\alpha \to 0$, the parameters can be estimated as $\gamma_1 \approx \alpha^2$ and $\gamma_2 \approx \alpha^2$. Note also Bob remain with the states

$$\left|\psi_0^{(01)\prime}\right\rangle = N_0^{(01)\prime}\left(a_0|0\rangle_1 + A_1^{(01)}(\alpha)\left(A_0^{(01)}(\alpha)A_0^{(01)}(\gamma_1)\right)^{-1}a_1|1\rangle_1\right), \qquad (44)$$

$$\left|\psi_1^{(01)\prime}\right\rangle = N_1^{(01)\prime}\left(a_0|0\rangle_1 + A_0^{(01)}(\alpha)\left(A_1^{(01)}(\alpha)A_1^{(01)}(\gamma_2)\right)^{-1}a_1|1\rangle_1\right), \qquad (45)$$

which he can distinguish and locally demodulate to increase the success probabilities in Eqs. (42) and (43). Here $N_0^{(01)\prime}$ and $N_1^{(01)\prime}$ are the corresponding normalization coefficients.

Consider another possibility for Bob to restore original unknown qubit from AM one by quantum swapping[41] with known state on example of demodulation of AM states with $k=0$ and $n=1$. Assume, that Bob has at his disposal the following states

$$N_0'\left(A_1^{(01)}A_0^{(01)-1}|01\rangle_{34} + |10\rangle_{34}\right), \qquad (46)$$

$$N_1'\left(A_0^{(01)}A_1^{(01)-1}|01\rangle_{34} + |10\rangle_{34}\right), \qquad (47)$$

where the normalization factors $N_0' = \left(1 + \left(\left|A_1^{(01)}\right|^2\left|A_0^{(01)}\right|^{-2} - 1\right)|a_1|^2\right)^{-0.5}$ and $N_1' = \left(1 + \left(\left|A_0^{(01)}\right|^2\left|A_1^{(01)}\right|^{-2} - 1\right)|a_1|^2\right)^{-0.5}$ are introduced. Bob mixes his AM qubits either $|\psi_{01}^{(01)}\rangle$ or $|\psi_{10}^{(01)}\rangle$ with the states in Eqs. (46) and (47), respectively, on balanced beam splitter[39] with subsequent registration of the measurement outcomes $|01\rangle, |10\rangle$ and reconstruct original unknown qubit in the base $\{|01\rangle, |10\rangle\}$. Then, summarizing all the probabilities, one can calculate the overall success probability

$$P_{t_1}^{(S)} = \frac{exp(-|\alpha|^2)}{1+\left(\left|A_0^{(01)}\right|^{-2}-1\right)|a_1|^2}\left(1 + \frac{|\alpha|^2(1-|\alpha|^2)^2}{|\alpha|^4+(1-|\alpha|^2)^2}\right), \qquad (48)$$

provided that Alice teleported the AM state in Eq. (24) and

$$P_{t_2}^{(S)} = \frac{exp(-|\alpha|^2)}{1+\left(\left|A_1^{(01)}\right|^{-2}-1\right)|a_1|^2}\left(1 + \frac{|\alpha|^2}{|\alpha|^4+(1-|\alpha|^2)^2}\right), \qquad (49)$$

in the case of the quantum teleportation of AM unknown qubit in Eq. (25). Note further state processing is no longer possible since all information about the states is lost in quantum swapping protocol. Superscript $S$ is responsible for the swapping operation.

**Taking into account higher-order measurement outcomes.** Let us consider a case of increase in the displacement amplitude $\alpha$. Then, the probability of other events only increases. Consider it on example of quantum teleportation of unknown qubit in Eq. (1). Now, we have to take into account contribution of the states $|\psi_{02}^{(01)}\rangle$ and $|\psi_{12}^{(01)}\rangle$, respectively, to the total success probability of quantum teleportation. It requires for Alice to send to Bob an additional bit (3 bits instead of 2). The calculation of overall success probability is based on Eqs. (40), (41). Consider the possibility of increasing the success probability on the example of interaction of the AM unknown qubit with a coherent state of large amplitude. Having received information from Alice that indicates that Bob has either a state $|\psi_{02}^{(01)}\rangle$ or $|\psi_{12}^{(01)}\rangle$, and knowing in advance the magnitude of the displacement used in the teleportation protocol, he mixes them with a coherent state with a predetermined amplitude as shown in section Methods. Bob uses large-amplitude coherent states $|0,-\varepsilon_3\rangle$ and $|0,-\varepsilon_4\rangle$, respectively, by analogy with how it is presented in Eqs. (92) and (94) with registration of those events which provide annihilation of amplitude-distorting coefficients. We can obtain quantities $\gamma_3$ and $\gamma_4$ from Eqs. (93) and (95). The overall success probabilities in Eqs. (42) and (43) are supplemented by additional terms to take into account contribution of the states $|\psi_{02}^{(01)}\rangle$ and $|\psi_{12}^{(01)}\rangle$ in restoring the original qubit. Following the procedure, one obtains the overall success probabilities

$$P_{t_1}^{(C)} = \frac{exp(-|\alpha|^2)}{1+\left(\left|A_0^{(01)}\right|^{-2}-1\right)|a_1|^2} \begin{pmatrix} 1 + exp(-|\gamma_1|^2)|\alpha|^2((1-|\gamma_1|^2)^2 + |\gamma_1|^2) + \\ exp(-|\gamma_3|^2) \frac{|\alpha|^4}{2!}((1-|\gamma_3|^2)^2 + |\gamma_3|^2) \end{pmatrix}, \quad (50)$$

$$P_{t_2}^{(C)} = \frac{exp(-|\alpha|^2)|\alpha|^2}{1+\left(\left|A_1^{(01)}\right|^{-2}-1\right)|a_1|^2} \begin{pmatrix} 1 + exp(-|\gamma_2|^2) |\gamma_2|^2/|\alpha|^2 + \\ exp(-|\gamma_4|^2) \frac{|\alpha|^2}{2!} |\gamma_4|^2 \end{pmatrix}, \quad (51)$$

where $\gamma_3 = -\alpha^2(1-|\alpha|^2)/(2-|\alpha|^2)$ and $\gamma_4 = -(1-|\alpha|^2)/(2-|\alpha|^2)$.

We can also make use of the quantum swapping method to get rid of the amplitude factor in the states $|\psi_{02}^{(01)}\rangle$ and $|\psi_{12}^{(01)}\rangle$ to increase the overall success probability. Bob prepares the similar states as in Eqs. (46), (47) with corresponding amplitude factors and mix them with AM states $|\psi_{02}^{(01)}\rangle$ and $|\psi_{12}^{(01)}\rangle$. It allows to Bob to demodulate the states with some probability which gives Bob the opportunity to increase the overall success probability

$$P_{t_1}^{(S)} = \frac{exp(-|\alpha|^2)}{1+\left(\left|A_0^{(01)}\right|^{-2}-1\right)|a_1|^2}\left(1 + \frac{|\alpha|^2(1-|\alpha|^2)^2}{|\alpha|^4+(1-|\alpha|^2)^2} + \frac{|\alpha|^4}{2!}\frac{(2-|\alpha|^2)^2}{|\alpha|^4+(2-|\alpha|^2)^2}\right), \quad (52)$$

for the initially AM unknown qubit in Eq. (24) and

$$P_{t_2}^{(S)} = \frac{exp(-|\alpha|^2)}{1+\left(\left|A_1^{(01)}\right|^{-2}-1\right)|a_1|^2}\left(1 + \frac{|\alpha|^2}{|\alpha|^4+(1-|\alpha|^2)^2} + \frac{|\alpha|^2}{2!}\frac{(2-|\alpha|^2)^2}{(1-|\alpha|^2)^2+(2-|\alpha|^2)^2}\right), \quad (53)$$

for the initially AM unknown qubit in Eq. (25).

Corresponding plots of the success probabilities $P_{t_1}^{(C)}$, $P_{t_2}^{(C)}$, $P_{t_1}^{(S)}$ and $P_{t_2}^{(S)}$ are shown in figures 4 and 5, respectively. These dependencies give better results than those presented in Figs. 3, especially with $\alpha$ growing. Bob's efforts also allows him to expand the range of absolute amplitude values $|a_1|$ for which success probability is greater of 0.5. The increase in the amplitude of the displacement $\alpha$ makes possible to improve efficiency of this protocol in terms of success probability to restore original qubit. The probability of success may exceed 0.5 in a wider range of values of $|a_1|$.

**Quantum teleportation performed exclusively by Alice.** We examined the implementation of the protocol of quantum teleportation of AM unknown qubit, where the preparation of the qubits is considered to be preparatory operation and it is not part of the protocol. This operation can be performed by a third party of the protocol (Victor) who then forwards AM qubits to Alice and finally check fidelity of the teleported qubit. Consideration of the implementation of the protocol in such interpretation makes sense. Indeed, when considering the implementation of protocol within framework of Bell states formalism, the initial unknown qubit is prepared with the help of spontaneous parametric down converter and probability of the event is tiny. Therefore, if we take into account the probability of the generation of an unknown qubit, then the protocol is realized very rarely. Nevertheless, it is possible to take into account the probability of transformation of unknown qubit into AM and involve the procedure in the protocol. If the preparation of AM qubit is assigned to Alice, then the following result from Eqs. (40) and (41) holds

$$P_t = q_k P_{t_1} + q_n P_{t_2}, \tag{54}$$

where $q_k$ and $q_n$ are the probabilities to create the AM states $|\varphi_{AMk}^{(kn)}\rangle$ and $|\varphi_{AMn}^{(kn)}\rangle$, respectively, from original unknown qubit.

The demodulation methods are also applicable to direct implementation of the protocol of quantum teleportation presented in Fig. 1(a) without prior amplitude modulation of unknown qubit. It is possible to choose the value of the displacement amplitude so as to ensure performance of the condition $A_m^{(kn)}(\alpha) = 1$. Corresponding state becomes original. Other AM states should be subject to amplitude demodulation. Calculations show that success probability is no longer dependent on the amplitude of the teleporting qubit and can take values larger of 0.5.

## Discussion

We have developed a new implementation of quantum teleportation of an unknown qubit which can be called DV-CV teleportation. DV-CV teleportation differs from already widely known DV[1] and CV[15-17] ones. DV-CV teleportation is based on the nonlinear effect when DV and CV states interact on the beam splitter. The mechanism of DV-CV interaction is different from measurement induced nonlinearity[10-14,20-22]. We have shown the DV-CV teleportation can be implemented with arbitrary settings. The choice of the setting parameters (amplitude $\beta$ of the hybrid state, transmittance $t$ and reflectance $r$ of the bam splitter) affects only amplitude-distorting factors in Eqs. (72), (73). We have shown that DV-CV teleportation is applicable to an unknown qubit in Eq. (77) with arbitrary numbers $k$ and $n$. So, the quantum teleportation of an unknown qubit beyond Bell states formalism can be recognized universal. Note the protocol is realized with an irreducible number of optical elements and hybrid entangled state in Eq. (78) can be realized using the same technique[27]. A simplified model of the DV-CV protocol also shows teleporting an unknown qubit with the correct selection of the parameters. The most vivid manifestation of the nonlinear effect associated with change of the displacement amplitude on the teleported qubit is observed in the case of different parity of the base states of original qubit. The developed protocol is free from the problems inherent to Bell states formalism in linear optics domain. All received qubits are identified unlike the Bell states measurement where only two Bell states give indistinguishable measurement outcomes. The proposed protocol is nearly deterministic as all measurement outcomes are identified. In this sense, the developed protocol recalls the CV teleportation of unknown qubit which is realized with help of two-mode squeezed vacuum. CV teleportation[15] is produced in deterministic manner but with restricted fidelity due to the fact that the quantum channel does not possess the maximum entanglement in the case of finite squeezing parameter. In the case of DV-CV teleportation, entangled hybridity formed from coherent components and single photon is used unlike two-mode squeezed vacuum. The problem of the DV-CV teleportation is reduced to finding, a way for Bob to demodulate the obtained qubits or the same to enhance fidelity of the teleported qubit.

Increase in the efficiency of quantum teleportation of an unknown qubit is an important task[19-22] which, in the standard formulation of the DV teleportation, is limited to 0.5 within the framework of linear optics methods[18]. Attempts to overcome this limitation in efficiency are associated with auxiliary photons, complex quantum channels and hyperentanglement which are difficult to implement in practice[10,11,19-22]. In our case, we are interested in finding a strategy to increase the success probability for Bob to restore the original qubit which automatically improves the teleportation fidelity. To do it, we have considered DV-CV protocol with an initial modulation of the unknown qubit and analyzed Bob's efforts to eliminate the amplitude-distorting factor in his qubits. Two methods to demodulate AM qubits are considered. It was shown that a strategy with prepared AM unknown qubits (by Victor) allows to increase the efficiency of the protocol. Highly unbalanced unknown qubits either with $|a_1| \ll |a_0|$ or $|a_0| \ll |a_1|$ can be teleported with fidelity close to one. Consideration of more effective strategies aimed at increase of the fidelity of the teleportation can be the basis for the further development of the DV-CV quantum teleportation.

Standard benchmark for certifying quantum teleportation consists in surpassing the maximum average fidelity between the teleported and the perfect states that can be achieved classically $2/3$ if Alice and Bob are connected via a classical channel and Alice makes a projective measurement. Theoretically, quantum teleportation in four-dimensional Hilbert space is deterministic with unit fidelity[1,4], but in practice, only two measurement outcomes are different, while the measurement outcomes of the two other Bell states are indistinguishable[18]. If we take into account both distinguishable and indistinguishable outcomes in the protocol of quantum teleportation in four-dimensional Hilbert space and average the fidelity over all possible input states, then we will get an average fidelity $2/3$ comparable to the classical. If we use success probabilities in Figures 4 and 5 to calculate the fidelities of the output state, then we will find that it exceeds $> 2/3$. But the results of success probabilities can be improved if we consider the contribution of other possible states which will also increase the fidelity of the output state. In addition, we considered the case HTBS varying only parameter $\beta$, thus reducing the number of possible scenarios. If we take into account arbitrary setting parameters, the number of possible strategies aimed at increasing the fidelity of the output qubit can only increase. It is worth noting that we did the calculations for small values $k$ and $n$ for unknown qubit in Eq. (77). The use of unknown qubits with other basic states can also lead to changes in the fidelity of output states. In addition, we proposed only two probabilistic methods for AM demodulation of unknown qubit. If we will find a way to determinately get rid of the known amplitude-distorting factors, then the teleported state will become perfect. Now, question of the maximum fidelity of the DV-CV protocol remains open since there are a lot of possible scenarios and strategies to run. Each of them gives its amplitude-damping coefficients that finally determine the output fidelity. All these studies require separate consideration.

## Methods

**Nonlinear mechanism of DV-CV interaction and quantum teleportation of unknown state on example of the state in Eq. (1).** When considering the interaction of CV and DV states, as a rule, analyze the case of mixing the states on highly transmissive beam splitter whose transmittance tends to unity ($t \to 1$). For example, the interaction of a coherent state with a very large amplitude $\beta$ (in theory $\beta \to \infty$) with any arbitrary state $|\varphi\rangle$ on HTBS leads to the displacement of the state on phase plane[29]

$$BS_{12}(|0,\beta\rangle_1 \otimes |\varphi\rangle_2) \approx |0,\beta\rangle_1 \otimes D_2(-\alpha)|\varphi\rangle_2. \quad (55)$$

where the notation $\otimes$ means tensor product and THE displacement amplitude can be taken $\alpha = \beta r$ in the limit case of $t = 1$. Here and throughout the entire paper it is assumed that the displacement amplitude $\alpha$ takes positive values $\alpha > 0$ as $\beta > 0$. Operator $D(\alpha)$ is the displacement one[42]

$$D(\alpha) = exp(\alpha a^+ - \alpha^* a), \quad (56)$$

with $\alpha$ being an amplitude of the displacement and $a$, $a^+$ are the bosonic annihilation and creation operators, over the number states. This extreme case has practical application as a reasonably efficient way to realize the displacement operator. It is natural to assume that the DV-CV interaction is not restricted to the limiting case and allows for one much more types of output states in the case of the arbitrary states input to the beam splitter with arbitrary parameters. Here, we consider a mathematical model of the DV-CV interaction on example of hybrid state in Eq. (2) with the qubit in Eq. (1) on an arbitrarily beam splitter (BS) in Eq. (3) as applied to the problem of quantum teleportation of an unknown qubit. Above, the case was considered in the simplified version.

Displaced number states will be widely used in the exact calculation
$$|n, \alpha\rangle = D(\alpha)|n\rangle. \tag{57}$$
In particular, the coherent state can be defined as result of action the displacement operator on vacuum $|\alpha\rangle \equiv |n, \alpha\rangle = D(\alpha)|0\rangle$. We are going to use the decomposition of the displaced number states in base of the number states[31]
$$|n, \alpha\rangle = F \sum_{m=0}^{\infty} c_{nm}(\alpha)|n\rangle, \tag{58}$$
where the overall multiplier $F(\alpha) = exp(-|\alpha|^2/2)$ is introduced. The amplitudes of the decomposition or matrix elements[31] are the following
$$c_{nm}(\alpha) = \frac{\alpha^{m-n}}{\sqrt{n!}\sqrt{m!}} \sum_{k=0}^{l} (-1)^k C_n^k |\alpha|^{2k} \prod_{k=0}^{n-1}(m-n+k+1), \tag{59}$$
where $C_n^k = n!/(k!(n-k)!)$ are the elements of the Bernoulli distribution and $\prod_{k=0}^{n-1}(m-n+k+1) = m(m-1)...(m-n+1)$ is the integer product. The matrix element $c_{nm}(\alpha)$ satisfy the normalization conditions[31]
$$\langle n, \alpha|k, \alpha\rangle = F^2 \sum_{m=0}^{\infty} c_{nm}^*(\alpha) c_{km}(\alpha) = \delta_{nk}, \tag{60}$$
where $\delta_{nk}$ is Kronecker delta[42]. For example, amplitudes for the coherent state $|\alpha\rangle$ and displaced single photon $|1, \alpha\rangle$ are given in Eqs. (10), (11).

The matrix elements $c_{nm}(\alpha)$ are the expressions involving a common factor proportional to $\alpha^{m-n}$ and polynomial of degree $l$ over variable $|\alpha|^2$ which is enclosed in parentheses. The polynomial in parentheses is invariant when changing the variable $\alpha \to \alpha exp(i\phi)$, where $\phi$ is an arbitrary phase. Therefore, only the factor $\alpha^{m-n}$ defines the behavior of the matrix elements under the change $\alpha \to \alpha exp(i\phi)$. Then, we have $\alpha^{m-n} \to \alpha^{m-n} exp(i(m-n)\phi)$ and the following property of the matrix elements
$$c_{nm}(-\alpha exp(i\phi)) = (-exp(i\phi))^{m-n} c_{nm}(\alpha), \tag{61}$$
being key for realization of the quantum teleportation of unknown state. If we take $\phi = 0$, then we have
$$c_{nm}(-\alpha) = (-1)^{m-n} c_{nm}(\alpha). \tag{62}$$
In particular, variation of the sign in the amplitude $\alpha \to -\alpha$ for coherent and displaced single photon state gives
$$c_{0n}(-\alpha) = (-1)^n c_{0n}(\alpha), \tag{63}$$
$$c_{1n}(-\alpha) = (-1)^{n-1} c_{1n}(\alpha). \tag{64}$$
Consider the action of the nonlinear mechanism on the example of the state in Eq. (1). Due to linearity of the beam splitter operator in Eq. (3), we have
$$BS_{12}(|\Psi\rangle_{134}|\varphi\rangle_2) = \frac{1}{\sqrt{2}} \left( BS_{12} \begin{pmatrix} |0, -\beta\rangle_1 (a_0|0\rangle_2 + a_1|1\rangle_2)|01\rangle_{34} + \\ |0, \beta\rangle_1 (a_0|0\rangle_2 + a_1|1\rangle_2)|10\rangle_{34} \end{pmatrix} \right) =$$
$$\frac{1}{\sqrt{2}} (BS_{12}(|0, -\beta\rangle_1 (a_0|0\rangle_2 + a_1|1\rangle_2))|01\rangle_{34} + BS_{12}(|0, \beta\rangle_1 (a_0|0\rangle_2 + a_1|1\rangle_2))|10\rangle_{34}), \tag{65}$$
Consider the operator's action $BS_{12}$ on states $|0, -\beta\rangle_1 (a_0|0\rangle_2 + a_1|1\rangle_2)|01\rangle_{34}$ and $|0, \beta\rangle_1 (a_0|0\rangle_2 + a_1|1\rangle_2)|10\rangle_{34}$, separately. We have the following chain of exact mathematical transformations
$$BS_{12}(|0, -\beta\rangle_1 (a_0|0\rangle_2 + a_1|1\rangle_2)|01\rangle_{34}) =$$
$$BS_{12}(D_1(-\beta)D_2(-\alpha)D_2(\alpha)|0\rangle_1 (a_0|0\rangle_2 + a_1|1\rangle_2)|01\rangle_{34}) =$$
$$BS_{12}(D_1(-\beta)D_2(-\alpha)BS_{12}^+ BS_{12}|0\rangle_1 D_2(\alpha)(a_0|0\rangle_2 + a_1|1\rangle_2)|01\rangle_{34}) =$$
$$D_1(-\beta/t)D_2(0)BS_{12}|0\rangle_1 (a_0|0, \alpha\rangle_2 + a_1|1, \alpha\rangle_2)|01\rangle_{34} =$$
$$F(\alpha)D_1(-\beta/t) \sum_{n=0}^{\infty} (a_0 c_{on}(\alpha) + a_1 c_{1n}(\alpha)) BS_{12}|0n\rangle_{12} |01\rangle_{34}$$
$$F(\alpha)D_1(-\beta/t) \sum_{n=0}^{\infty} \frac{(a_0 c_{on}(\alpha) + a_1 c_{1n}(\alpha))}{\sqrt{n!}} (ra_1^+ + ta_2^+)^n |00\rangle_{12} |01\rangle_{34}, \tag{66}$$
for the first term and
$$BS_{12}(|0, \beta\rangle_1 (a_0|0\rangle_2 + a_1|1\rangle_2)|10\rangle_{34}) =$$
$$BS_{12}(D_1(\beta)D_2(\alpha)D_2(-\alpha)|0\rangle_1 (a_0|0\rangle_2 + a_1|1\rangle_2)|10\rangle_{34}) =$$
$$BS_{12}(D_1(\beta)D_2(\alpha)BS_{12}^+ BS_{12}|0\rangle_1 D_2(-\alpha)(a_0|0\rangle_2 + a_1|1\rangle_2)|10\rangle_{34}) =$$

$$D_1(\beta/t)D_2(0)BS_{12}|0\rangle_1(a_0|0,-\alpha\rangle_2 + a_1|1,-\alpha\rangle_2)|10\rangle_{34} =$$
$$F(\alpha)D_1(\beta/t)\sum_{n=0}^{\infty}(a_0c_{on}(-\alpha) + a_1c_{1n}(-\alpha))BS_{12}|0n\rangle_{12}|10\rangle_{34}$$
$$F(\alpha)D_1(\beta/t)\sum_{n=0}^{\infty}(-1)^n(a_0c_{on}(\alpha) - a_1c_{1n}(\alpha))BS_{12}|0n\rangle_{12}|10\rangle_{34}$$
$$F(\alpha)D_1(\beta/t)\sum_{n=0}^{\infty}(-1)^n\frac{(a_0c_{on}(\alpha)-a_1c_{1n}(\alpha))}{\sqrt{n!}}(ra_1^+ + ta_2^+)^n|00\rangle_{12}|10\rangle_{34}, \qquad (67)$$

where $D^+(\beta), BS_{12}^+$ are Hermitian conjugate ones and $D(\beta)D^+(\beta) = I$, $BS_{12}BS_{12}^+ = I$ with identical operator $I$. An exact condition $\alpha = \beta r/t = const$ is kept. Note that in the case of the HTBS, the amplitude of the coherent state should be chosen to be much larger ($\beta \to \infty$) to ensure that this condition is performed with reflectance tending to zero $r \to 0$. Here, we do not impose such strict restrictions on the amplitude of the coherent state and on the values of the parameters of the BS. The equations (66) and (67) include the operator sum $ra_1^+ + ta_2^+$ in power $n$. Expanding the sum and collecting the terms corresponding to $k$ photons in the second mode $|k\rangle_2$, one obtains

$$BS_{12}(|\Psi\rangle_{134}|\varphi\rangle_2) = \frac{F(\alpha)}{\sqrt{2}}\sum_{k=0}^{\infty}t^k(|\Psi_k\rangle_1|01\rangle_{34} + |\Psi'_k\rangle_1|10\rangle_{34})|k\rangle_2, \qquad (68)$$

where

$$|\Psi_k\rangle_1 = \sum_{n=0}^{\infty}r^n\sqrt{C_{n+k}^k}(a_0c_{0n+k}(\alpha) + a_1c_{1n+k}(\alpha))|n,-\beta/t\rangle_1, \qquad (69)$$

$$|\Psi'_k\rangle_1 = \sum_{n=0}^{\infty}(-1)^{n+k}r^n\sqrt{C_{n+k}^k}(a_0c_{0n+k}(\alpha) - a_1c_{1n+k}(\alpha))|n,\beta/t\rangle_1. \qquad (70)$$

Now, if Alice will measure $p$ photons in the first mode and $k$ photons in the second mode, then Bob obtains the following conditional state under arbitrary setting parameters (the amplitude $\beta$ of the state in Eq. (2) and the transmittance $t$ and the reflectance $r$ of the BS in Eq. (3), respectively)

$$\left|\varphi_{pk}^{(01)}\right\rangle_2 = N_{pk}^{(01)}\left(a_0A_{pk}^{(01)}|01\rangle_2 + a_1B_{pk}^{(01)}|10\rangle_2\right), \qquad (71)$$

provided that Alice has supplied him with information about her measurement outcomes so that he can perform the initial transformation operations over the output qubit: Z-transformation in the case of $p + k$ being odd number (Z-transformation is not used in the case of $p + k$ being even number) replaced by action of Hadamard transformation in Eqs. (7a) and (7b) regardless of the parity of the sum $p + k$. Here, Bob's state acquires additional real known factors $A_{pk}^{(01)}$ and $B_{pk}^{(01)}$ that we denote both with superscripts $(01)$ (to distinguish the factors from those that may appear in the case of quantum teleportation of an unknown qubit with basic states different from vacuum and single photons) and subscripts $(pk)$ being measurement outcomes. The factors and the normalization multiplier $N_{pk}^{(01)}$ are given by

$$A_{pk}^{(01)} = \sum_{n=0}^{\infty}r^n\left(\sqrt{C_{n+k}^k}c_{0n+k}(\alpha)c_{np}(\beta/t) - r\sqrt{C_{n+k+1}^k}c_{0n+k+1}(\alpha)c_{n+1p}(\beta/t)\right), \qquad (72)$$

$$B_{pk}^{(01)} = \sum_{n=0}^{\infty}r^n\left(\sqrt{C_{n+k}^k}c_{1n+k}(\alpha)c_{np}(\beta/t) - r\sqrt{C_{n+k+1}^k}c_{1n+k+1}(\alpha)c_{n+1p}(\beta/t)\right), \qquad (73)$$

$$N_{pk}^{(01)} = \left(|a_0|^2\left|A_{pk}^{(01)}\right|^2 + |a_1|^2\left|B_{pk}^{(01)}\right|^2\right)^{-1/2}. \qquad (74)$$

In general form, the success probability $P_{pk}^{(01)}$ for Bob to deal with the state in Eq. (71) is given by

$$P_{pk}^{(01)} = t^{2k}F^2(\alpha)F^2(\beta/t)\left(|a_0|^2\left|A_{pk}^{(01)}\right|^2 + |a_1|^2\left|B_{pk}^{(01)}\right|^2\right). \qquad (75)$$

Using the relations in Eqs. (58) and (60), it is possible to directly check the success probabilities are normalized

$$\sum_{k=0}^{\infty}\sum_{p=0}^{\infty}P_{pk}^{(01)} = 1. \qquad (76)$$

Note that the displaced states in Eq. (57) are intermediate in the calculation. These states do not remain in the final expressions and leave their trace exclusively in the amplitude-distorting factors of the teleported state. The property of the displaced states to change the sign of the displacement amplitudes, depending on the parity of the displaced state, is a cornerstone in the quantum teleportation of unknown qubit with help of the state in Eq. (2). Since the parity of the displaced states $|0, -\alpha\rangle$ and $|1, -\alpha\rangle$ is different, this leads to the fact that the amplitudes of decomposition $c_{1n}(\alpha)$ acquire additional sign " $-$ " on compared with $c_{0n}(\alpha)$ as $\left(a_0 c_{on}(-\alpha) + a_1 c_{1n}(-\alpha) = (-1)^n (a_0 c_{on}(\alpha) - a_1 c_{1n}(\alpha))\right)$. The nonlinear mechanism different from measurement-induced nonlinearity is similar to action of the controlled sign change gate. We only note that if we consider the teleportation of an unknown qubit, for example, $|\varphi^{(02)}\rangle_2 = a_0|0\rangle_2 + a_1|2\rangle_2$, where another base state $|2\rangle$ is used, then the nonlinear effect is not observed for the quantum channel in Eq. (2) since the parity of the basic states is the same.

After the initial operations, Bob stays with the qubit in Eq. (71) which contains an additional real known factor $B_{pk}^{(01)}/A_{pk}^{(01)}$ that distorts it. If the setting parameters are chosen in such a way that the condition $B_{pk}^{(01)}/A_{pk}^{(01)} = 1$ for the certain measurement outcomes $(pk)$ is fulfilled, then Bob gets the initial qubit in Eq. (1) but only in basis $\{|01\rangle, |10\rangle\}$. It is natural to assume that for other measurement outcomes this condition may not be performed and Bob should get rid of the unwanted factor to increase the fidelity of the teleportation. Let us use "fairly loosely" the term amplitude-modulated (AM) qubit with respect to the state in Eq. (71) to distinguish it from the perfect in Eq. (1). Also "loosely", we are going to use the term demodulation to procedure of getting rid of the amplitude factor $B_{pk}^{(01)}/A_{pk}^{(01)}$ and restoring the original qubit by Bob. Naturally, not all measurement outcomes can be taken into account by Alice and Bob. The probabilities of some events in Eq. (75) may significantly exceed the remaining ones, which allows participants of the protocol to neglect "minor events". Therefore, in this case, we should only talk about the nearly-deterministic quantum teleportation of the unknown qubit. Then, Alice and Bob's task is, by variation of the setting parameters, to develop a strategy that would allow Alice to teleport an unknown qubit to Bob with the highest possible success probability with a minimal amount of classical information sent to Bob. Moreover, this strategy should help Bob to demodulate the qubit in Eq. (71) with the greatest possible success probability. The terms introduced here are used throughout the work.

Despite the fact that developed approach is valid for arbitrary values of the amplitude of the hybrid state $\beta$ and parameters $t$ and $r$ of the BS, it makes sense to analyze strategies in the case of an arbitrary value $\beta$ and the parameters of the HTBS $t \to 1$ and $r \to 0$. In this case, the expressions (72) and (73) are simplified and become $A_{pk}^{(01)} = c_{0k}(\alpha)c_{0p}(\beta)$ and $B_{pk}^{(01)} = c_{1k}(\alpha)c_{0p}(\beta)$. These expressions were used to derive the coefficient $A_m^{(01)}$ in Eq. (14) and probability $P_m^{(01)}$ in Eq. (18), where summation over the subscript $p$ was performed. Note only the fact that we used other notations $A_m^{(01)}$ in Eq. (14) and $P_m^{(01)}$ in Eq. (18) to distinguish them from the notations used in the case of arbitrary used parameters $\beta$ and $t$.

**Quantum teleportation of unknown state of the state $a_0|k\rangle_2 + a_1|n\rangle_2$.** In the previous section, we considered the possibility of quantum teleportation of the state in Eq. (1). Here we are going to generalize this result on the case of quantum teleportation of unknown state

$$|\varphi^{(kn)}\rangle_2 = a_0|k\rangle_2 + a_1|n\rangle_2, \qquad (77)$$

satisfying the normalization condition $|a_0|^2 + |a_1|^2 = 1$ and superscript $(kn)$ is introduced to denote basic qubit states. The same DV-CV nonlinear interaction mechanism described in the previous section will be used. To fulfill the basic condition in Eq. (61) of the mechanism we must already use another quantum channel

$$|\Psi_\phi\rangle_{134} = (|0,-\beta\rangle_1|01\rangle_{34} + |0,exp(i\phi)\beta\rangle_1|10\rangle_{34})/\sqrt{2}, \tag{78}$$

where the amplitude $\beta$ may acquire arbitrary values and one of the coherent components of the hybrid state acquires an additional phase factor $\phi$. Let us consider the case of the mixing of these states on HTBS with $t \to 1$ and $r \to 0$

$$|\Delta^{(kn)}\rangle_{1234} = BS_{12}\left(|\Psi_\phi\rangle_{134}|\varphi^{(kn)}\rangle_2\right) = 1/\sqrt{2}$$
$$\left(BS_{12}\left(|0,-\beta\rangle_1|\varphi^{(kn)}\rangle_2\right)|01\rangle_{34} + BS_{12}\left(|0,exp(i\phi)\beta\rangle_1|\varphi^{(kn)}\rangle_2\right)|10\rangle_{34}\right). \tag{79}$$

In order to implement the DV-CV nonlinear mechanism, we must impose the following condition on the phase $\phi$

$$(-exp(i\phi))^{k-n} = -1. \tag{80}$$

Then, the output can be written as

$$|\Delta^{(kn)}\rangle_{1234} = \frac{F(\beta)}{\sqrt{2}(1+exp(i\phi))}$$
$$\sum_{m=0}^{\infty} \frac{c_{km}}{N_m^{(kn)}} \left( \begin{array}{c} \frac{|\Psi_{+\phi}\rangle_1}{N_{+\phi}} \left(Zexp(i\phi/2)R_Z(\phi)\right)^{m-k} H|\Psi_m^{(kn)}\rangle_{34} + \\ \frac{|\Psi_{-\phi}\rangle_1 exp(-i\phi)}{N_{-\phi}} \left(Zexp(i\phi/2)R_Z(\phi)\right)^{m-k+1} H|\Psi_m^{(kn)}\rangle_{34} \end{array} \right) |m\rangle_2, \tag{81}$$

where we introduce normalized amplitude-distorting state $|\Psi_m^{(kn)}\rangle$ as

$$|\Psi_m^{(kn)}\rangle = N_m^{(kn)}\begin{bmatrix} a_0 \\ A_m^{(kn)} a_1 \end{bmatrix}, \tag{82}$$

with the normalization factor

$$N_m^{(kn)} = \left(1 + \left(\left|A_m^{(kn)}\right|^2 - 1\right)|a_1|^2\right)^{-0.5}, \tag{83}$$

where the amplitude-distorting multiplier is the following

$$A_m^{(kn)} = \frac{c_{nm}}{c_{km}}. \tag{84}$$

The superposition states are defined by

$$|\Psi_{+\phi}\rangle_1 = N_{+\phi}(|0,-\beta\rangle_1 + exp(-i\phi)|0,exp(i\phi)\beta\rangle_1), \tag{85}$$
$$|\Psi_{-\phi}\rangle_1 = N_{-\phi}(|0,-\beta\rangle_1 - |0,exp(i\phi)\beta\rangle_1), \tag{86}$$

where $N_{+\phi}$ and $N_{-\phi}$ are the corresponding normalization factors. It is worth noting that the states in Eqs. (85) and (86) are transformed into even/odd SCS in the case of $\phi = 0$

$$|even\rangle = N_+(|0,-\beta\rangle + |0,\beta\rangle), \tag{87}$$
$$|odd\rangle = N_-(|0,-\beta\rangle - |0,\beta\rangle), \tag{88}$$

where the factors $N_\pm = \left(2(1 \pm exp(-2|\beta|^2))\right)^{-1/2}$ are the normalization parameters. Note that states in Eqs. (85) and (86) can be approximated as

$$|\Psi_{+\phi}\rangle_1 \approx |0\rangle_1, \tag{89}$$
$$|\Psi_{-\phi}\rangle_1 \approx |1\rangle_1, \tag{90}$$

in the case of $\beta < 1$. $Z$ is the corresponding Pauli matrix and $R_Z(\phi)$ is a rotation operation about $Z$ axis. Consider the unknown qubit in Eq. (77) with an odd difference $n - k$ of basis elements. Then, to satisfy the condition in Eq. (80), we must put $\phi = 0$. Suppose the numbers $k$ and $n$ are connected by the relation $n = 4l + 2 + k$ between each other with $l = 0,1,2, .....$ Then, we have to chose $\phi = \pi/2$ to provide performance of the condition in Eq. (80). Finally, if we consider $n = 4l + k$, then the phase is equal to $\phi = \pi/4l$ with $l = 0,1,2, .....$

At the next stage, Alice must determine the number of photons in the second mode and distinguish the outcomes of states $|\Psi_{+\phi}\rangle$ and $|\Psi_{-\phi}\rangle$ from each other in first mode. Hadamard operation $H$ (Eqs. (7a) and (7b)) and rotation $R_Z(\phi)$ are implemented by Bob on single qubit[40] after receiving a message from Alice about her measurement outcomes. Finally, Bob obtains one of AM states in Eq. (82) with probability

$$P_m^{(kn)} = \frac{F^2|c_{km}|^2}{|1+exp(i\phi)|N_m^{(kn)2}}\left(\frac{1}{N_{+\phi}^2} + \frac{1}{N_{-\phi}^2}\right), \tag{91}$$

satisfying the normalization condition $\sum_{m=0}^{\infty} P_m^{(kn)} = 1$. It should be noted that the states in Eqs. (85) and (86) are orthogonal to each other in the case of $n-k$ being odd. If the difference $n-k$ becomes even then the states in Eqs. (85), (86) become approximately orthogonal in the domain of $\beta < 1$.

**Demodulation of AM state by interaction of it with strong coherent state.** Consider one of the possibilities for Bob to demodulate AM states either $|\psi_{01}^{(01)}\rangle$ or $|\psi_{10}^{(01)}\rangle$ by its interaction with strong coherent state on HTBS. For example, use the state $|\psi_{01}^{(01)}\rangle_{23}$ occupying modes 2 and 3. The coherent state $|0,-\varepsilon_1\rangle_1$ with real amplitude $\varepsilon_1 > 0$ occupies first mode. Modes 1 and 3 are mixed on HTBS described by Eq. (3). In the case, such interaction results in displacement of the target state $|\psi_{01}^{(01)}\rangle_{23}$ by the quantity $\gamma_1 = \varepsilon_1 r$. Then, we have

$$BS_{13}\left(|0,-\varepsilon_1\rangle_1 |\psi_{01}^{(01)}\rangle_{23}\right) FN_{01}^{(01)\prime}|0,-\varepsilon_1\rangle_1$$
$$\begin{pmatrix} c_{10}(\gamma_1)\left(a_0|0\rangle_2 + A_1^{(01)}A_0^{(01)-1}(c_{00}(\gamma_1)/c_{10}(\gamma_1))a_1|1\rangle_2\right)|0\rangle_3 + \\ c_{11}(\gamma_1)\left(a_0|0\rangle_2 + A_1^{(01)}A_0^{(01)-1}(c_{01}(\gamma_1)/c_{11}(\gamma_1))a_1|1\rangle_2\right)|1\rangle_3 \end{pmatrix}. \tag{92}$$

Here, we take into account only the first two members of the superposition whose probabilities are maximal in the case of $\gamma_1 \ll 1$. The contributions of higher order terms are close to zero so they can be neglected in the superposition. If we adopt the condition

$$A_1^{(01)}A_0^{(01)-1}(c_{01}(\gamma_1)/c_{11}(\gamma_1)) = 1, \tag{93}$$

then Bob obtains the original unknown qubit in Eq. (1) provided that he registered the single photon in the third mode. If he fixed vacuum, then Bob obtains AM state in Eq. (44). We can estimate the value of the amplitude $\gamma_1$ from equation (93). Using the Eqs. (92) and (40), we can finally obtain the success probability $P_{t_1}^{(C)}$ in Eq. (42)).

The considered approach is applicable to demodulation of the state $|\psi_{10}^{(01)}\rangle_{23}$ with help of coherent state $|0,-\varepsilon_2\rangle_1$ with real amplitude $\varepsilon_2 > 0$ interacting with the state on HTBS. Indeed, we have

$$BS_{13}\left(|0,-\varepsilon_2\rangle_1 |\psi_{10}^{(01)}\rangle_{23}\right) FN_{10}^{(01)\prime}|0,-\varepsilon_2\rangle_1$$
$$\begin{pmatrix} c_{10}(\gamma_2)\left(a_0|0\rangle_2 + A_0^{(01)}A_1^{(01)-1}(c_{00}(\gamma_2)/c_{10}(\gamma_2))a_1|1\rangle_2\right)|0\rangle_3 + \\ c_{11}(\gamma_2)\left(a_0|0\rangle_2 + A_0^{(01)}A_1^{(01)-1}(c_{01}(\gamma_2)/c_{11}(\gamma_2))a_1|1\rangle_2\right)|1\rangle_3 \end{pmatrix}. \tag{94}$$

in regime $\gamma_2 \ll 1$. If we impose the following condition

$$A_1^{(01)}A_0^{(01)-1}(c_{00}(\gamma_2)/c_{10}(\gamma_2)) = 1, \tag{95}$$

then Bob has at his disposal the original qubit (1) provided that he registered vacuum in the third mode. If the result of his measurement is a single photon, then he obtains the AM state in Eq. (45). Thus, value of $\gamma_2$ can be obtained from Eq. (95). Using the Eq. (94), we can calculate the overall success probability in Eq. (43) to demodulate AM unknown qubti.

**Acknowledgement**

The work was supported by Act 211 Government of the Russian Federation, contract № 02.A03.21.0011.

**Author Contributions**

Authors S.A.P. contributed to the conception of the mechanism and development of mathematical apparatus. S.A.P. executed the numerical simulations. Both (S.A.P. and J.K.) discussed, wrote and reviewed the manuscript.

**Additional Information**

**Competing financial and/or non-financial interests:** We declare that the authors have no competing (financial) interests as defined by Nature Research, or other (non-financial) interests that might be perceived to influence the results and/or discussion reported in this paper.

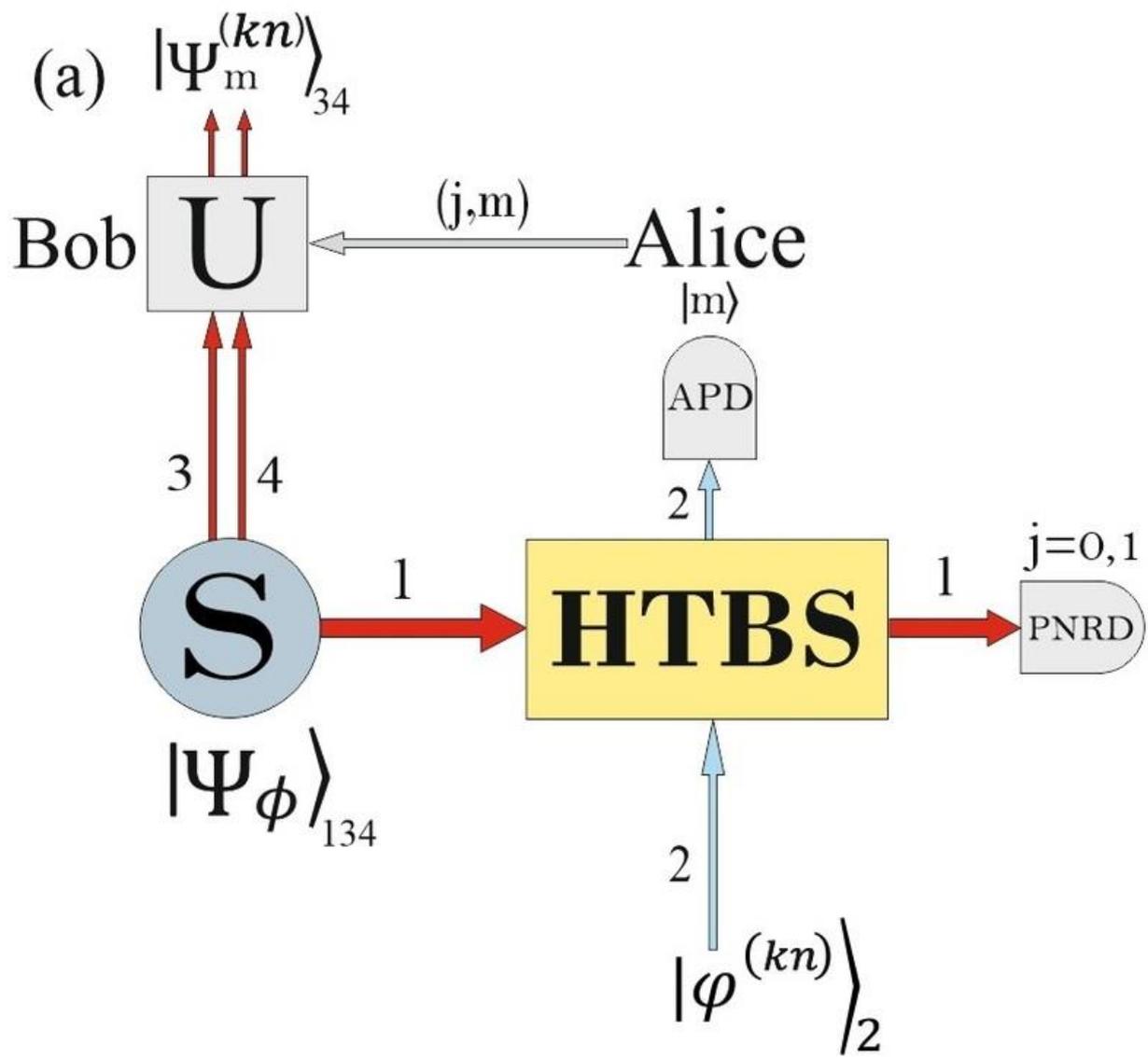

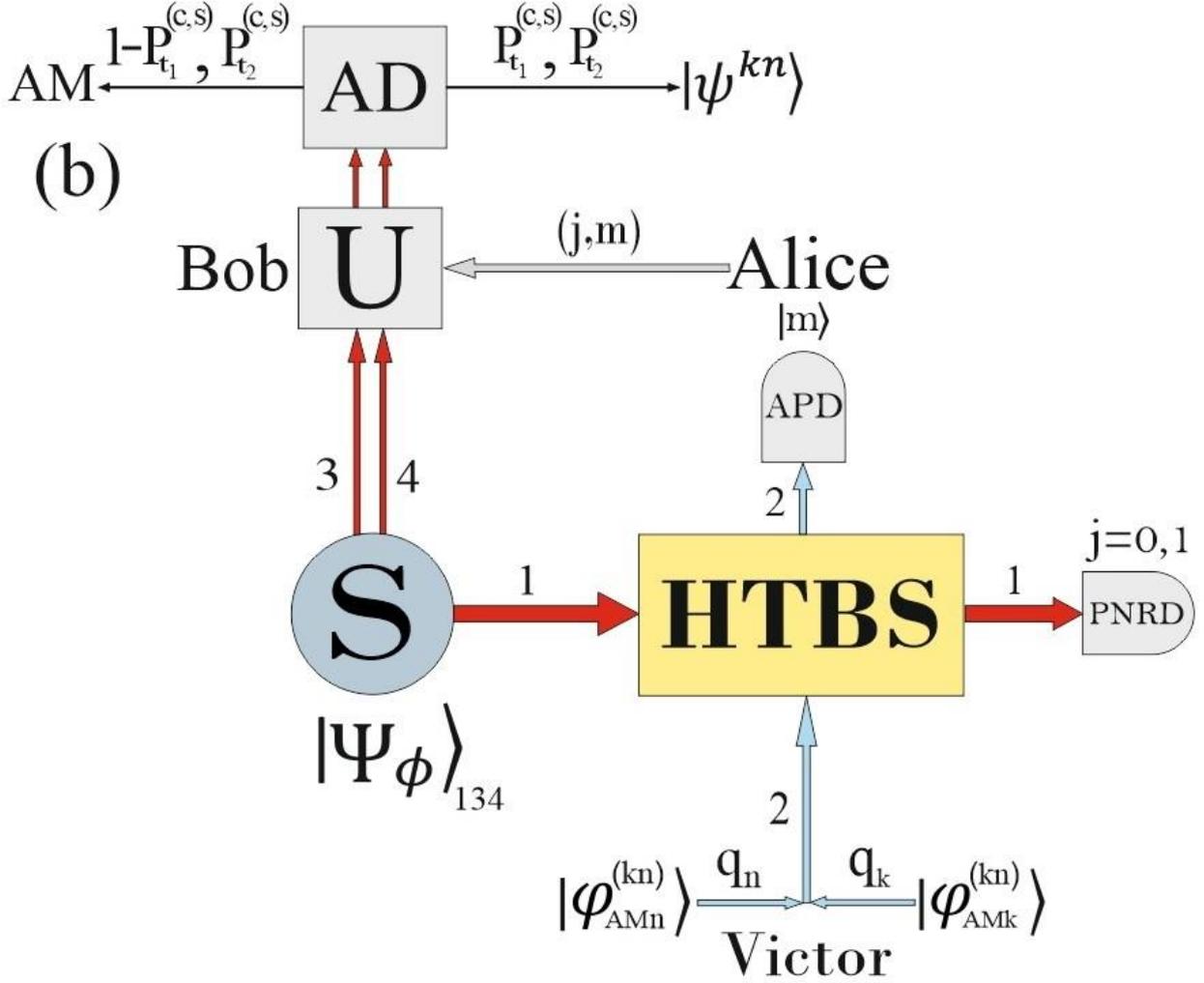

**Figure 1.** (a) Unknown qubit $|\varphi^{(kn)}\rangle_2$ interacts with coherent components of the entangled state $|\Psi_\phi\rangle_{134}$ (designation $S$) on HTBS with subsequent measurement in both modes. In general case, measurement may be performed with APD and photon number resolving detector (PNRD) and results in four outcomes in the case of $\alpha < 1$. Bob performs unitary transformation $U = H(Zexp(i\phi/2)R_Z(\phi))^{m-k}$ on his single photon conditioned by classical information $(j, m)$ from Alice and obtains AM qubits $|\Psi_m^{(kn)}\rangle$. (b) Victor modulates unknown qubit and sends to Alice AM qubits either $|\varphi_{AMk}^{(kn)}\rangle$ in Eq. (24) or $|\varphi_{AMn}^{(kn)}\rangle$ in Eq. (25), in general case, with corresponding probabilities $q_k$ and $q_n$ which may take values 0 and 1. This scheme includes the Bob's efforts to demodulate received qubits (AD means amplitude demodulation). Finally, Bob obtains original qubit with success probability either $P_{t_1}^{(C,S)}$ in Eqs. (50), (52) or $P_{t_2}^{(C,S)}$ in Eqs. (51), (53) depending on Bob's strategy to demodulate obtained qubits. Quantities $1 - P_{t_1}^{(C,S)}$ and $1 - P_{t_2}^{(C,S)}$ are the probabilities for Bob to stay with AM qubits with known amplitude-distorting factors.

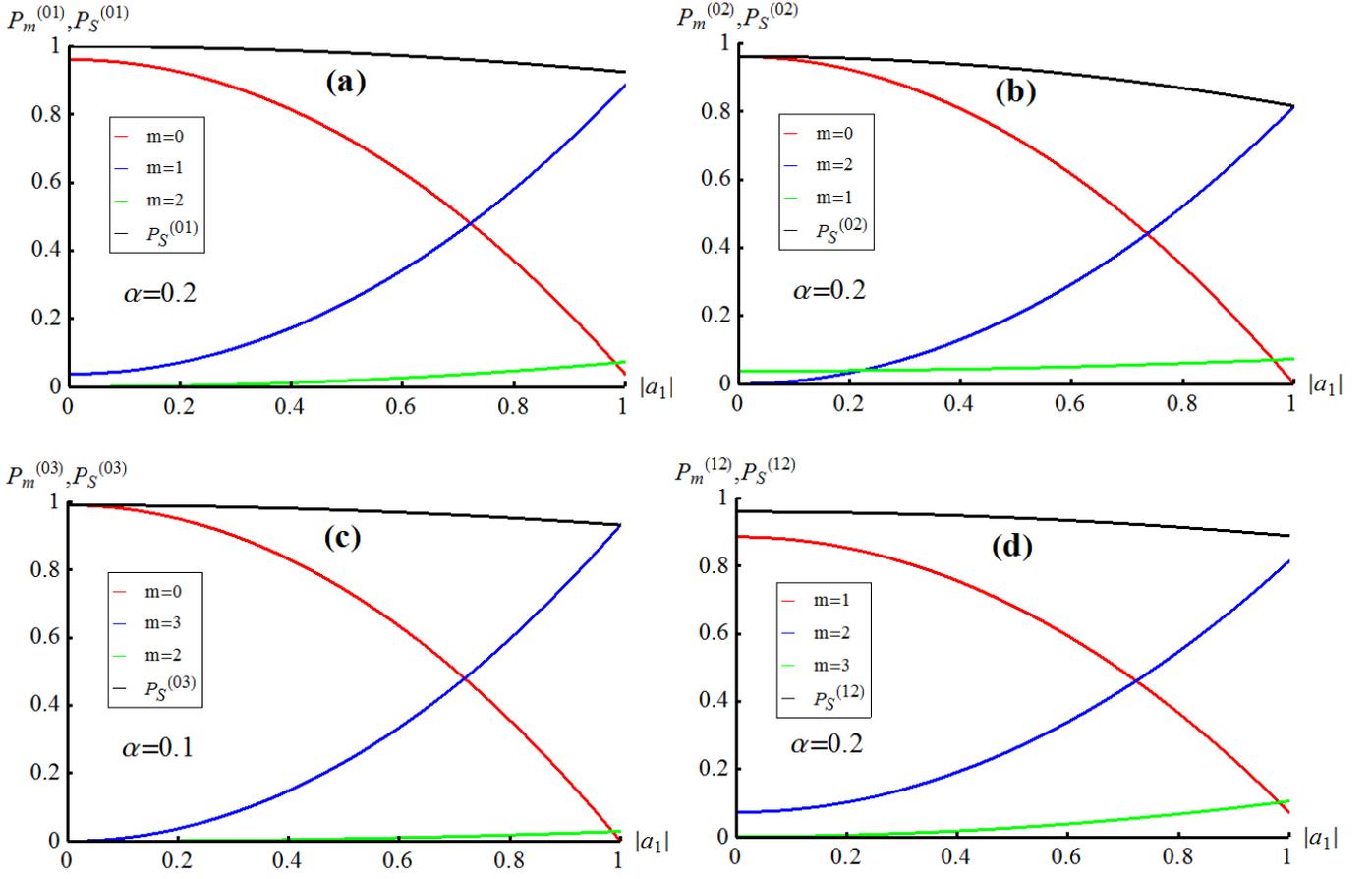

**Figure 2.** The success probability (**a**) $P_m^{(01)}$, $P_S^{(01)} = P_0^{(01)} + P_1^{(01)}$ (Eq. (18)), (**b**) $P_m^{(02)}$, $P_S^{(02)} = P_0^{(02)} + P_2^{(02)}$, (**c**) $P_m^{(03)}$, $P_S^{(03)} = P_0^{(03)} + P_3^{(03)}$ and (**d**) $P_m^{(12)}$, $P_S^{(12)} = P_1^{(12)} + P_2^{(12)}$ (Eq. (92)) as functions of $|a_1|$ for different values of $\alpha$ and $m$ as indicated in each subfigure. The condition $P_S^{(kn)} \approx 1$ (Eq. (21)) is performed.

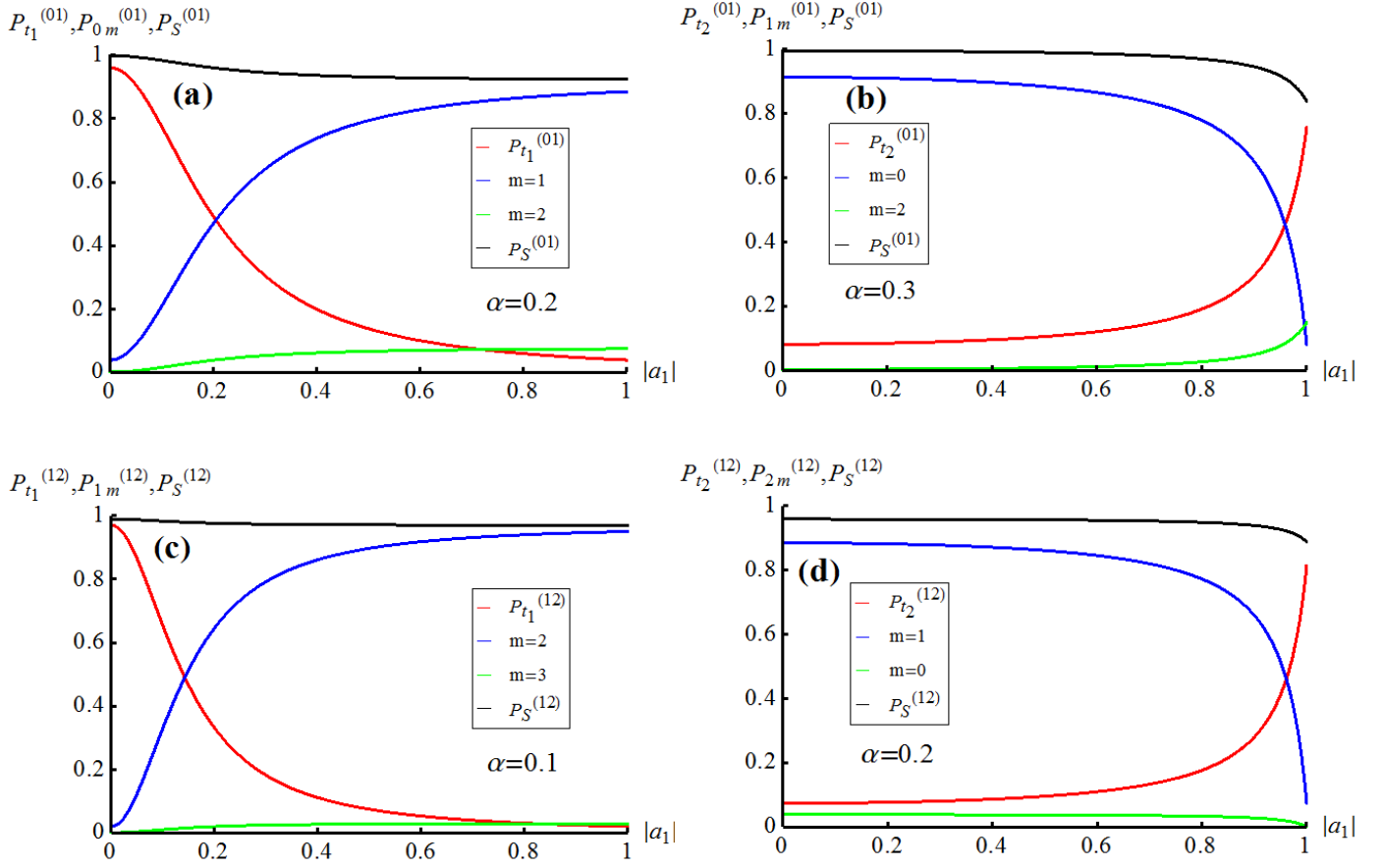

**Figure 3.** The success probabilities (**a**) $P_{t_1}^{(01)}$, $P_{01}^{(01)}$, $P_{02}^{(01)}$, $P_S^{(01)} = P_{t_1}^{(01)} + P_{01}^{(01)}$, (**b**) $P_{t_2}^{(01)}$, $P_{10}^{(01)}$, $P_{12}^{(01)}$, $P_S^{(01)} = P_{t_2}^{(01)} + P_{10}^{(01)}$, (**c**) $P_{t_1}^{(12)}$, $P_{12}^{(12)}$, $P_{13}^{(12)}$, $P_S^{(12)} = P_{t_1}^{(12)} + P_{12}^{(12)}$ and (**d**) $P_{t_2}^{(12)}$, $P_{21}^{(12)}$, $P_{20}^{(12)}$, $P_S^{(12)} = P_{t_2}^{(12)} + P_{21}^{(12)}$ (Eqs. (29), (30) and Eqs. (35), (36)) as functions of $|a_1|$ for different values of $\alpha$ and $m$ as indicated in each subfigure. As can be seen from the plots the conditions $P_S^{(01)} \approx 1$, $P_S^{(12)} \approx 1$ (Eqs. (37), (38)) are performed.

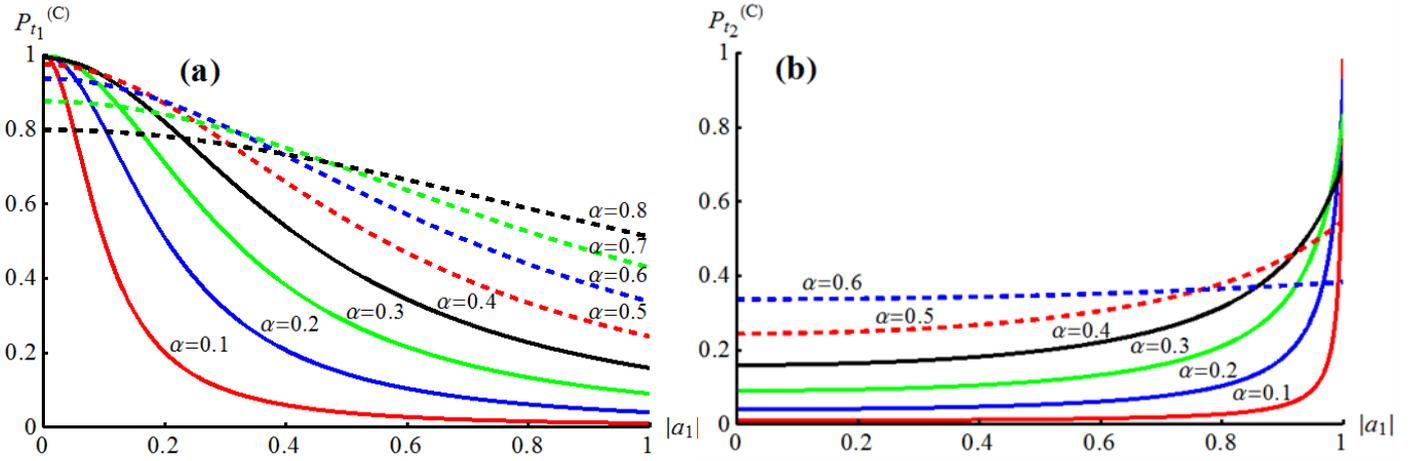

**Figure 4.** Plots of the success probabilities (a) $P_{t_1}^{(C)}$ (Eq. (50)) and (b) $P_{t_2}^{(C)}$ (Eq. (51)) to teleport AM unknown qubit either $|\varphi_{AM0}^{(01)}\rangle$ in Eq. (24) or $|\varphi_{AM1}^{(01)}\rangle$ in Eq. (25) taking into account contribution of the AM states $|\psi_{02}^{(01)}\rangle$ and $|\psi_{12}^{(01)}\rangle$ and possibility for Bob to demodulate AM states by means of their interaction with coherent state of large amplitude (Eqs. (92), (94)) in dependence on $|a_1|$ for different values of the displacement amplitude $\alpha$ indicated on the curves.

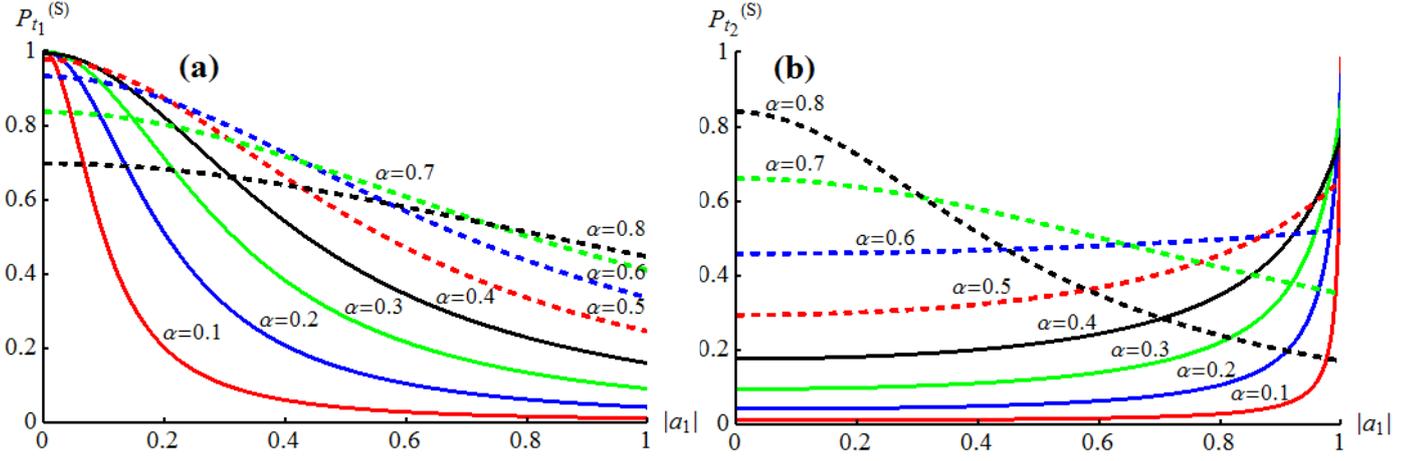

**Figure 5.** Plots of the success probabilities (**a**) $P_{t_1}^{(S)}$ (Eq. (52)) and (**b**) $P_{t_2}^{(S)}$ (Eq. (53)) to teleport AM unknown qubit either $|\varphi_{AM0}^{(01)}\rangle$ in Eq. (24) or $|\varphi_{AM1}^{(01)}\rangle$ in Eq. (25) taking into account contribution of the AM states $|\psi_{02}^{(01)}\rangle$ and $|\psi_{12}^{(01)}\rangle$ and possibility for Bob to demodulate AM states by means of quantum swapping with the states in Eqs. (44), (45) in dependence on $|a_1|$ for different values of the displacement amplitude $\alpha$ indicated on the curves.